\documentclass[acmsmall]{acmart}
\usepackage{graphicx}
\usepackage{xspace}
\usepackage{listings}
\usepackage{enumitem}
\usepackage{diagbox}
\usepackage{booktabs}
\usepackage{pgfplots}
\usepgfplotslibrary{statistics}
\pgfplotsset{compat=1.18}
\usepackage{subcaption}
\usepackage{wrapfig}
\usepackage{adjustbox}
\usepackage[most]{tcolorbox}
\usepackage{tabularx}
\usepackage{multirow}
\usepackage{makecell}
\usepackage[table]{xcolor}
\usepackage{hyperref}

\usepackage[textsize=scriptsize]{todonotes}

\newif\ifclean
\cleanfalse 

\ifclean
    
\else
    
\fi

\definecolor{codegreen}{rgb}{0,0.6,0}
\definecolor{codegray}{rgb}{0.5,0.5,0.5}
\definecolor{codepurple}{rgb}{0.58,0,0.82}
\definecolor{backcolour}{rgb}{0.95,0.95,0.92}

\usepackage{ifthen}
\usepackage{xcolor}

\def\BibTeX{{\rm B\kern-.05em{\sc i\kern-.025em b}\kern-.08em
    T\kern-.1667em\lower.7ex\hbox{E}\kern-.125emX}}

\definecolor{commentred}{HTML}{0000FF}

\newboolean{showcomments}
\setboolean{showcomments}{false}

\ifthenelse{\boolean{showcomments}}
  {
   \newcommand{\mrnote}[2]{%
     {\color{commentred}%
        \fbox{\bfseries\sffamily #1}%
        {\small$\blacktriangleright$\sffamily\itshape #2 $\blacktriangleleft$}%
     }%
   }
   \newcommand{\mrm}[1]{{\color{commentred}#1}}
  }
  {
   \newcommand{\mrnote}[2]{#2}
   \newcommand{\mrm}[1]{#1}
  }

\newcommand{\mr}[2]{\mrnote{Response to #1 }{#2}}

\lstdefinestyle{trajectory}{
  basicstyle=\ttfamily\small,
  columns=fullflexible,
  breaklines=true,
  breakatwhitespace=false,
  keepspaces=true,
  showstringspaces=false,
  frame=single,
  numbers=none,
  numberstyle=\tiny,
  xleftmargin=1.6em,
  framexleftmargin=1.2em,
  aboveskip=0.6\baselineskip,
  belowskip=0.6\baselineskip,
  escapeinside={(*@}{@*)},
  moredelim=**[is][\color{red}\ttfamily]{@}{@},
  moredelim=**[is][\color{red}\ttfamily]{<<}{>>},
  moredelim=**[is][\color{blue}\ttfamily]{[[}{]]}
}

\usepackage{listings}
\usepackage{xcolor}

\lstdefinestyle{compact}{
    basicstyle=\footnotesize\ttfamily, 
    frame=single,                      
    framesep=2pt,                      
    aboveskip=5pt,                     
    belowskip=5pt,                     
    numbers=left,                      
    numberstyle=\tiny\color{gray},     
    numbersep=4pt,                     
    captionpos=b,                      
    breaklines=true,                   
    backgroundcolor=\color{white},     
    keywordstyle=\color{blue},
    commentstyle=\color{green!40!black},
    stringstyle=\color{orange},
    showstringspaces=false
}

\definecolor{codegreen}{rgb}{0,0.6,0}
\definecolor{codegray}{rgb}{0.5,0.5,0.5}
\definecolor{codepurple}{rgb}{0.58,0,0.82}
\definecolor{shallowred}{rgb}{1,0.8,0.8}
\definecolor{verylightgray}{rgb}{0.97, 0.97,0.97}

\lstdefinestyle{mystyle}{
    language=Java, 
    commentstyle=\color{codegreen},
    keywordstyle=\color{blue},
    stringstyle=\color{codepurple},
basicstyle=\fontfamily{zi4}\selectfont\scriptsize,
    breakatwhitespace=false,
    breaklines=true,              
    captionpos=b,                    
    keepspaces=true,                 
    numbers=left,                    
    numbersep=5pt,                  
    showspaces=false,                
    showstringspaces=false,
    frame=single,                    
    xleftmargin=0.03\linewidth,
    xrightmargin=0.03\linewidth,
    framexleftmargin=0mm,            
    framexrightmargin=0mm,           
    framextopmargin=0mm,             
    framexbottommargin=0mm,    
    showtabs=false,                  
    tabsize=2,
    escapeinside={(*@}{@*)},
}

\lstdefinestyle{trajectoryy}{
    basicstyle=\fontfamily{zi4}\selectfont\scriptsize,
    columns=fullflexible,
    breaklines=true,
    breakatwhitespace=false,
    keepspaces=true,
    showstringspaces=false,
    showspaces=false,
    showtabs=false,
    tabsize=2,
    frame=none,
    numbers=left,
    captionpos=b,
    numbersep=5pt,
    numberstyle=\tiny,
    framexleftmargin=0.05mm,
    framexrightmargin=0.05mm,
    framextopmargin=0mm,
    framexbottommargin=0mm,
    aboveskip=0pt,
    belowskip=0pt,
    escapeinside={(*@}{@*)},
    moredelim=**[is][\color{red}\ttfamily]{@}{@},
    moredelim=**[is][\color{red}\ttfamily]{<<}{>>},
    moredelim=**[is][\color{blue}\ttfamily]{[[}{]]}
}

\newtcblisting{TrajectoryListing}[1][]{%
listing only,
  breakable=false,
  colframe=black!30,
  boxrule=0pt,
  arc=0.1pt,
  colback=white,
  left=0mm,right=0mm,top=0mm,bottom=0mm,
  fonttitle=\footnotesize\bfseries, 
  colbacktitle=black!30,
  coltitle=white,
  toptitle=0mm,                  
  bottomtitle=0mm,               
  lefttitle=1.5mm,
  righttitle=1.5mm,
  listing options={style=trajectoryy},
  #1
}

\NewDocumentCommand{\mynote}{+O{}+m}{%
  \begingroup
  \tcbset{
    noteshift/.store in=\mynote@shift,
    noteshift=1.5cm,
  }
  \begin{tcolorbox}[
    enhanced,
    boxrule=0.45pt,
    colback=yellow!7,
    colframe=yellow!45!black,
    left=1mm,   
    right=1mm,  
    top=0.4mm,      
    bottom=0.4mm,   
    before skip=4pt, after skip=6pt,
    arc=1mm,
    fonttitle=\bfseries,
    coltitle=black,
    attach boxed title to top left={xshift=3mm,yshift*=-2mm},
    boxed title style={
      colback=yellow!25,
      colframe=yellow!45!black,
      boxrule=0.45pt,
      arc=1mm,
      left=1mm,right=1mm,top=0.4mm,bottom=0.4mm,
    },
    overlay={\node[right] (mynotenode) at ([xshift=-\mynote@shift]frame.west) {} ;},
    #1
  ]
  #2
  \end{tcolorbox}
  \endgroup
}

\setcopyright{cc}
\setcctype{by}
\acmDOI{10.1145/3832100}
\acmYear{2026}
\acmJournal{PACMSE}
\acmVolume{3}
\acmNumber{ISSTA}
\acmArticle{ISSTA009}
\acmMonth{10}
\acmSubmissionID{issta26main-p86-p}
\received{2026-01-30}
\received[accepted]{2026-06-25}

\begin{document}

\title{Sifting the Noise: A Comparative Study of LLM Agents in Vulnerability False Positive Filtering}

\author{Yunpeng Xiong}
\orcid{0009-0003-4330-4549}
\affiliation{%
  \institution{Monash University}
  \city{Melbourne}
  \country{Australia}
}
\email{nemo.xiong@monash.edu}

\author{Ting Zhang}
\orcid{0000-0002-6001-1372}
\affiliation{%
  \institution{Monash University}
  \city{Melbourne}
  \country{Australia}
}
\email{ting.zhang@monash.edu}

\authornote{Ting Zhang is the corresponding author.}

\begin{abstract}
Static Application Security Testing (SAST) tools are essential for identifying software vulnerabilities, but they often produce a high volume of False Positives (FPs), imposing a substantial manual triage burden on developers.
Recent advances in Large Language Model (LLM) agents offer a promising direction by enabling iterative reasoning, tool use, and environment interaction to refine SAST alerts.
However, the comparative effectiveness of different LLM-based agent architectures for FP filtering remains poorly understood.

In this paper, we present a comparative study of three state-of-the-art LLM-based agent frameworks, i.e., \textsc{Aider}, \textsc{OpenHands}, and \textsc{SWE-agent}, for vulnerability FP filtering.
We evaluate these frameworks using the vulnerabilities from the OWASP Benchmark and real-world open-source Java projects.
We further conduct a focused post-cutoff C/C++ study using the strongest configuration to test contamination-free generalization and isolate key agentic capabilities.
The experimental results show that LLM-based agents can remove the majority of SAST noise, reducing an initial FP detection rate of over 92\% on the OWASP Benchmark to as low as 6.3\% in the best configuration.
On a real-world Java dataset, the best configuration of LLM-based agents can achieve an FP identification rate of up to 93.3\% involving CodeQL alerts.
However, the benefits of agents are strongly backbone- and CWE-dependent: agentic frameworks significantly outperform vanilla prompting for stronger models such as Claude Sonnet 4 and GPT-5, but yield limited or inconsistent gains for weaker backbones.
On the post-cutoff OSS-Fuzz dataset, \textsc{SWE-agent} with Claude Sonnet 4 identifies 95.5\% of FPs while maintaining 95.5\% precision, compared with a 36.4\% FP identification rate for vanilla prompting.
Moreover, aggressive FP reduction can come at the cost of suppressing true vulnerabilities, highlighting important trade-offs.
Finally, we observe large disparities in computational cost across agent frameworks.

Overall, our study demonstrates that LLM-based agents are a powerful but non-uniform solution for SAST FP filtering, and that their practical deployment requires careful consideration of agent design, backbone model choice, vulnerability category, and operational cost.
\end{abstract}



\maketitle

\section{Introduction}

Recent years have seen a surge of software vulnerabilities, driven by the rapid growth of software systems and their increasing complexity~\cite{nistnvd,nvdCWEOverTime,kharkar2022reduce}.
The sheer number of reported vulnerabilities poses challenges to software security, as development teams struggle to detect, prioritize, and remediate security issues in a timely manner~\cite{dissanayake2022patch,li2017securitypatches,cheng2025vercation,cheng2026fixseeker}.

\textit{\textbf{Problem and its importance.}}
Static Application Security Testing (SAST) tools have been widely adopted in both industry and open-source communities to identify vulnerabilities early in the software development lifecycle~\cite{owaspsast,sadowski2015tricorder,calcagno2015moving,li2023comparison}.
These tools analyze source code without execution and can be seamlessly integrated into IDEs and CI/CD pipelines~\cite{owaspsast,sonarqube,codeql}.
However, SAST tools are well known for producing a high volume of FPs~\cite{kang2022detecting, aloraini2019empirical,guo2023mitigating,lenarduzzi2023comparision,zhang2026titanca}, i.e., warnings that resemble vulnerabilities syntactically but are not exploitable in practice.
This noise increases the manual triage burden on developers and often leads to alert fatigue, reduced trust in automated security tools, or even blanket suppression of warnings ~\cite{aloraini2019empirical,guo2023mitigating,hu2025empirical}.

In our empirical evaluation on the OWASP Benchmark~\cite{owaspbenchmark}, we observe that this problem is particularly severe:
When aggregating findings across multiple mainstream SAST tools, the vast majority of non-vulnerable cases are flagged as suspicious by at least one scanner.
Moreover, many FPs are shared across tools, indicating correlated semantic blind spots rather than isolated tool-specific errors.
These observations suggest that simply combining or tuning existing SAST tools is insufficient to make their outputs actionable at scale.

\textit{\textbf{Existing techniques.}}
To mitigate the FP problem, prior research has explored learning-based approaches that classify or prioritize static warnings~\cite{guo2023mitigating,kang2022detecting}.
For example, Transformer-based models have been \mrm{used} to distinguish FPs from true vulnerabilities using learned representations of code and warnings~\cite{kharkar2022reduce}.
More recently, LLMs have been applied as semantic inspectors for static analysis results~\cite{wen2024llm4sa,du2025llm4pfa}.
The state-of-the-art approach, LLM4SA~\cite{wen2024llm4sa}, demonstrates that prompting LLMs with SAST warnings and relevant code context can improve precision.

Despite these advances, existing LLM-based techniques primarily rely on \emph{static prompting}, where the model is used as a one-shot classifier that produces a verdict in a single pass~\cite{wen2024llm4sa,du2025llm4pfa}.
Such approaches cannot actively explore codebases, gather additional evidence, or iteratively refine responses, in contrast to agentic frameworks that can interact with environments and tools~\cite{yao2022react,yang2024sweagent,wang2025openhands}.
However, these abilities are often essential for accurately assessing whether a reported vulnerability is real or an FP, especially in the presence of complex control flow or object-level semantics\cite{li2024interpvd,lenarduzzi2023comparision,wen2024llm4sa}.

\textit{\textbf{Our insights.}}
LLM-based agents have brought considerable breakthroughs in recent years by extending vanilla LLMs with iterative reasoning, tool use, and environment interaction~\cite{yao2022react,yang2024sweagent,wang2025openhands}.
In software engineering tasks, agent-based approaches have shown strong effectiveness in areas such as automatic program repair~\cite{yu2025patchagent,bouzenia2025repairagent}, agile effort estimation~\cite{bui2025llm}, and code generation~\cite{lin2025soen}.
Unlike static prompting, these agents operate in multi-step \textit{perceive--reason--act} loops, enabling them to inspect code, navigate repositories, and validate intermediate assumptions~\cite{yao2022react,yang2024sweagent,wang2025openhands}.

These capabilities naturally align with how human auditors triage SAST warnings, suggesting that LLM-based agents may be well-suited for vulnerability FP filtering~\cite{aloraini2019empirical,guo2023mitigating,wen2024llm4sa}.
However, it remains unclear whether agentic frameworks consistently outperform vanilla LLM prompting in this task, how different agent designs compare with each other, and what trade-offs they introduce in terms of accuracy, safety, and computational cost~\cite{wen2024llm4sa,yao2022react,yang2024sweagent,wang2025openhands}.

To fill this gap, we present a comparative study of three state-of-the-art LLM agent frameworks, i.e., \textsc{Aider}~\cite{aider}, \textsc{OpenHands}~\cite{wang2025openhands}, and \textsc{SWE-agent}~\cite{yang2024sweagent}, for filtering FP produced by SAST tools.
We evaluate these frameworks with Claude Sonnet 4, DeepSeek Chat, and GPT-5 under a unified task definition: given a static alert and codebase access, determine whether the report corresponds to a real vulnerability or a false alarm.
\mrm{Our full cross-model and cross-agent evaluation covers the OWASP Benchmark~\cite{owaspbenchmark} and real-world Java alerts from Vul4J~\cite{bui2022vul4j}, using Claude Sonnet 4, DeepSeek Chat, and GPT-5.
For this comparative evaluation, we standardize prompts, constrain external tooling, and measure FP filtering performance, true-vulnerability suppression risk, and operational overhead.
To address data-leakage and generalization concerns, we further conduct a focused post-cutoff C/C++ study on OSS-Fuzz~\cite{ossfuzz} using the strongest configuration (\textsc{SWE-agent} with Claude Sonnet 4) together with targeted ablations and stronger non-agentic baselines.}

Our results show that agentic approaches \mrm{substantially reduce SAST noise, but benefits depend on the backbone and weakness category.}
On OWASP Benchmark, the best configuration reduces the remaining FPR to single digits, with residual errors concentrated in hard weakness families such as cryptography- and policy-oriented CWEs.
Across the three models, agentic Claude and GPT match or outperform vanilla zero-shot prompting, whereas DeepSeek performs best without agents.
\mrm{On real-world CodeQL findings, agents remain effective.
Some frameworks are more aggressive at removing FPs, while others are more conservative to avoid suppressing true vulnerabilities.
The focused OSS-Fuzz study further shows that the strongest configuration generalizes to post-cutoff C/C++ alerts, with gains driven mainly by cross-file navigation and multi-turn interaction rather than larger static context.
Finally, computational overhead varies widely across frameworks, revealing a practical cost-effectiveness frontier for deploying agentic triage.}

In summary, this work makes three contributions:
\begin{itemize}[leftmargin=1em,label=$\star$,nosep]
    \item \textbf{Dimension.} This paper bridges recent advances in LLM-based autonomous agents~\cite{yao2022react,yang2024sweagent,wang2025openhands} with a long-standing and practically critical software security problem, i.e., FP filtering in SAST~\cite{guo2023mitigating,kang2022detecting}.
    While prior work has explored LLMs as one-shot semantic classifiers~\cite{wen2024llm4sa,du2025llm4pfa}, our study is the first to systematically examine whether agentic reasoning, tool use, and environment interaction can meaningfully improve the actionability of SAST results at scale, both on benchmarks and real-world codebases.
    \item \textbf{Comparative Study.} We conduct the first comprehensive comparative evaluation of three state-of-the-art LLM-based agent frameworks, i.e., \textsc{Aider}~\cite{aider}, \textsc{OpenHands}~\cite{wang2025openhands}, and \textsc{SWE-agent}~\cite{yang2024sweagent}, for vulnerability FP filtering. \mrm{The full cross-model and cross-agent comparison covers the OWASP Benchmark (v1.2) and real-world Java alerts from Vul4J~\cite{bui2022vul4j}, using Claude Sonnet 4, DeepSeek Chat, and GPT-5 as backbone models. We further conduct a focused post-cutoff C/C++ study on OSS-Fuzz~\cite{ossfuzz} using the strongest configuration to evaluate contamination-free generalization, stronger non-agentic baselines, and capability ablations.}
    \item \textbf{Empirical Insights and Practical Guidelines.}
    Our results show that LLM-based agents can remove the majority of SAST FPs, reducing an initial FP detection rate of over 92\% to 6.3\% on OWASP Benchmark in the best configuration. On a sample (n=50) of the Vul4J dataset, the agents show a maximum of 93.3\% FP identification rate when given alerts generated by CodeQL.
    However, we find that the benefit of shifting to an agentic framework is backbone- and CWE-dependent: agentic reasoning improves FP filtering for stronger models, but yields limited or even negative gains for others.
    Residual FPs are concentrated in policy- and cryptography-related CWEs, while injection-style vulnerabilities are filtered reliably.
    We further reveal trade-offs between aggressive FP removal, true vulnerability suppression, and operational cost, providing concrete guidance on when and how LLM-based agents should be deployed in practice.
\end{itemize}

\section{Background and Motivation}

\subsection{Static Application Security Testing Tools}
Static Application Security Testing (SAST) tools automatically scan application source code, bytecode, or binaries to identify security vulnerabilities such as injection flaws, XSS, and buffer overflows. These tools facilitate white-box analysis by integrating into IDEs and CI/CD pipelines to detect flaws early in the development lifecycle \cite{owaspsast, lenarduzzi2023comparision}. By analyzing code without execution, SAST tools transform source text into intermediate representations to identify security violations \cite{wen2024llm4sa, li2023comparison}.

\noindent\textbf{Query-based SAST Tools.} Query-based SAST tools, exemplified by CodeQL \cite{codeql}, Joern \cite{joern}, and Semgrep \cite{bennett2024semgrep}, parse a codebase into a relational database representing the program structure. This database encapsulates information such as abstract syntax trees, control flow graphs, and program dependence graphs~\cite{lenarduzzi2023comparision, li2024evaluating, li2024interpvd}. 
Vulnerability patterns are codified as domain-specific queries, such as CodeQL's object-oriented query language or Joern's graph traversals, which search the database for data-flow paths connecting sources to sinks.

\noindent\textbf{Limitations and FPs.}
Despite their sophistication, the SAST tools are severely compromised by the trade-off between precision and scalability \cite{lenarduzzi2023comparision}. 
Industry leaders like Google, Microsoft, and Meta deploy these tools at scale to maintain code safety \cite{sadowski2015tricorder, calcagno2015moving, li2024evaluating}. 
In practice, they often sacrifice precision (context sensitivity and path sensitivity) in favor of efficiency to maintain scalability over millions of lines of code \cite{wen2024llm4sa}.

Moreover, these tools report a high rate of FPs \cite{li2024evaluating}. 
FPs from SAST tools cause "alert fatigue," resulting in developers losing trust in automated security tooling \cite{hu2025empirical}. 
When faced with thousands of warnings, manual triage becomes prohibitively expensive. 
Consequently, developers frequently resort to suppression mechanisms to silence reports. However, evidence shows that 50.8\% of suppressions are "useless": they do not actively suppress any warning due to code changes, yet they may even unintentionally mask future legitimate vulnerabilities \cite{hu2025empirical}.



\subsection{Motivating Example: A Concrete False Positive}
In this section, we first present one example \texttt{BenchmarkTest00171.java} (Listing \ref{lst:example_of_a_FP}) from the OWASP Benchmark to show when traditional SAST tools produce an FP on benign code.

\begin{TrajectoryListing}
// ... (imports)
public void doPost(HttpServletRequest request, HttpServletResponse response) {
    // 1. Taint Source: 'param' receives user input
    String param = request.getHeader("BenchmarkTest00171");
    param = java.net.URLDecoder.decode(param, "UTF-8");

    // 2. Data Flow Obfuscation via HashMap
    java.util.HashMap map40534 = new java.util.HashMap();
    map40534.put("keyA-40534", "a_Value"); // Safe value
    map40534.put("keyB-40534", param);     // Tainted value

    // 3. Retrieval: 'bar' retrieves the SAFE value from keyA
    String bar = (String) map40534.get("keyA-40534");

    // 4. Sink: Command execution using 'bar'
    String cmd = "echo " + bar;
    Runtime.getRuntime().exec(cmd);

    // ...
}
\end{TrajectoryListing}
\captionof{lstlisting}{An FP example from OWASP benchmark (BenchmarkTest00171.java)}
\label{lst:example_of_a_FP}

In this scenario, the variable \texttt{param} is indeed tainted by user input.
The code stores this tainted value in a HashMap under \texttt{keyB}, but subsequently retrieves a safe, hardcoded value (\texttt{"a\_Value"}) from \texttt{keyA} to assign to the variable \texttt{bar}. 
Finally, \texttt{bar} is used in a system command execution.

\begin{itemize}[left=0pt, labelsep=0.5em, itemsep=0pt]
    \item \textbf{CodeQL} reports that the "shell command line depends on a user-provided value," failing to distinguish the specific key-value dependencies within the map operations.
    \item \textbf{Semgrep} warns that "Untrusted input might be injected into a command," relying on coarse-grained taint tracking that observes data entering the map and data leaving the map, without precise sensitivity to the string keys.
\end{itemize}

These tools produce FPs as they lack precise object or path sensitivity to HashMaps, failing to track whether the retrieved value \texttt{bar} is semantically different from the tainted \texttt{param}.
This example highlights a fundamental limitation of existing SAST tools: their inability to precisely reason about object-level semantics, motivating the need for more context-aware filtering approaches.
A human auditor can readily deduce that the tainted data flow is severed. 
This suggests that an LLM-based agent with sufficient semantic reasoning capabilities may be able to reach the same conclusion.
As shown by our experimental results (Section~\ref{rq1:result}), existing SAST tools produce a high rate of FPs, indicating a strong need for more effective filtering mechanisms.
\subsection{LLMs for Vulnerability Analysis: From Oracles to Agents}
\noindent\textbf{LLMs as Passive Semantic Oracles.} LLMs have demonstrated effectiveness in function-level vulnerability detection by interpreting complex control flows and variable relationships \cite{zhou2024large, gnieciak2025large,li2025out}.
In the context of static analysis, they often act as semantic oracles to post-process warnings \cite{wadhwa2024core}. For instance, LLM4SA demonstrated that feeding SAST findings and code context into an LLM can improve precision by distinguishing true vulnerabilities from FPs \cite{wen2024llm4sa}. Other approaches have coupled LLM reasoning with program analysis to recover richer context, such as inferring taint specifications \cite{li2025iris} or constructing vulnerability-focused slices via code property graphs \cite{lekssays2025llmxcpg}.

Despite these advancements, most prior methods treat LLMs as one-shot classifiers operating on fixed, pre-extracted snippets. As probabilistic text generators, these vanilla LLM approaches lack the structural capability to verify their own outputs or actively navigate multi-file codebases, which limits their performance in rigorous security auditing.

\noindent\textbf{The Emergence of Autonomous Agents.} To address these limitations, recent research has shifted toward autonomous agents that operate in a \textit{Perceive-Reason-Act-Observe} loop \cite{yao2022react}. Unlike single-pass LLMs, agentic frameworks iteratively collect evidence and use external tools to refine hypotheses before reaching a decision \cite{yang2024sweagent, wang2025openhands}. While agents have shown promising results in repository-level tasks such as program repair \cite{yu2025patchagent, bouzenia2025repairagent}, their efficacy in filtering FPs produced by SAST tools remains unexamined.

This gap motivates our study. We investigate the effectiveness of LLM-based agents for vulnerability FP triage through a comparative analysis of three representative frameworks. Our evaluation spans multiple backbone models, utilizing both a curated benchmark and real-world projects to provide actionable guidance on the role of agentic triage in software security.

\section{Study Design}

\begin{figure}
    \centering
    \includegraphics[width=\linewidth]{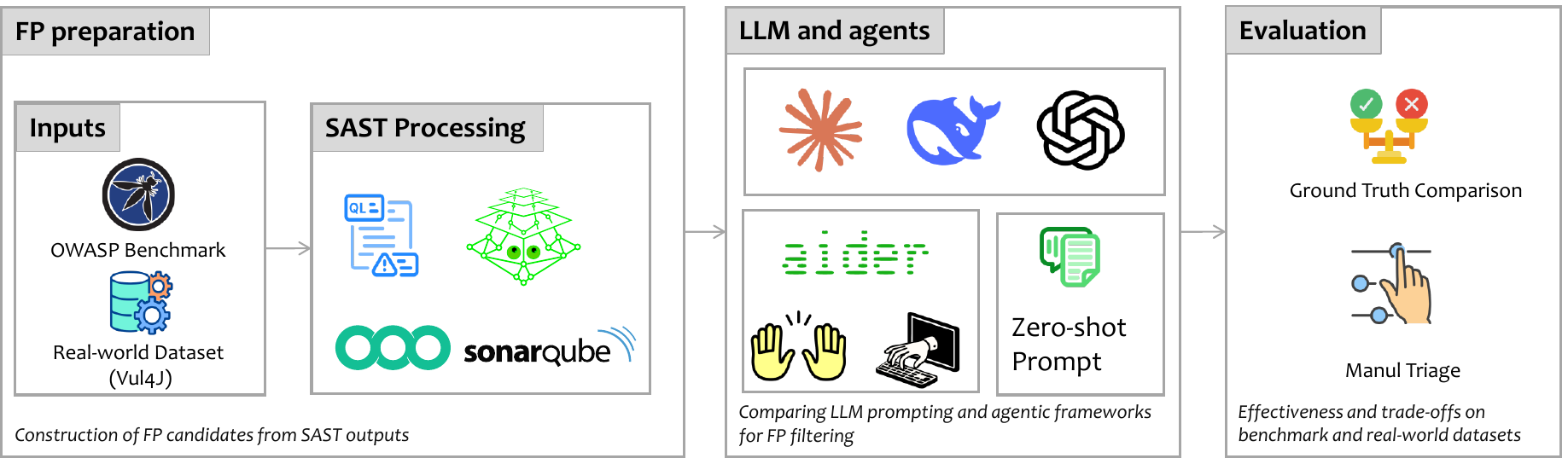}
    \caption{Overview of our work}
    \label{fig:experiment_design}
\end{figure}

\subsection{Research Questions}
Figure~\ref{fig:experiment_design} shows the overview of our study. 
Specifically, in this study, we aim to answer the following research questions (RQs):

\begin{itemize}[leftmargin=1em,nosep]
    \item \textbf{RQ1:} How effective are different LLM-based agent frameworks in filtering FPs generated by SAST tools?
    \item \textbf{RQ2:} How effective are the LLM-based agent frameworks in identifying FPs in real-world \mrm{Java} scenarios?
    \mrm{\item \textbf{RQ3:} What agentic capabilities drive FP filtering performance, and do the gains generalize to a contamination-free, C/C++ setting?}
    \item \textbf{RQ\mrm{4}:} What are the key success drivers and recurring failure modes of LLM-based agents in FP identification?
\end{itemize}

\subsection{Methodology for RQ1} 
\label{rq1:methodology}



\begin{table}[h]
\caption{Data statistics of the OWASP Benchmark for Java, version 1.2}
\centering
\small
\begin{tabular}{lccc}
\toprule
\textbf{Vulnerability Area} & \textbf{Instances} & \textbf{Positive} & \textbf{Negative} \\
\midrule
CWE-78 (Command Injection) & 251 (9.16\%) & 126 (4.60\%) & 125 (4.56\%) \\
CWE-327 (Weak Cryptography) & 246 (8.98\%) & 130 (4.74\%) & 116 (4.23\%) \\
CWE-328 (Weak Hashing) & 236 (8.61\%) & 129 (4.71\%) & 107 (3.91\%) \\
CWE-90 (LDAP Injection) & 59 (2.15\%) & 27 (0.99\%) & 32 (1.17\%) \\
CWE-22 (Path Traversal) & 268 (9.78\%) & 133 (4.85\%) & 135 (4.93\%) \\
CWE-614 (Secure Cookie Flag) & 67 (2.45\%) & 36 (1.31\%) & 31 (1.13\%) \\
CWE-89 (SQL Injection) & 504 (18.39\%) & 272 (9.93\%) & 232 (8.47\%) \\
CWE-501 (Trust Boundary Violation) & 126 (4.60\%) & 83 (3.03\%) & 43 (1.57\%) \\
CWE-330 (Weak Randomness) & 493 (17.99\%) & 218 (7.96\%) & 275 (10.04\%) \\
CWE-643 (XPATH Injection) & 35 (1.28\%) & 15 (0.55\%) & 20 (0.73\%) \\
CWE-79 (Cross-Site Scripting / XSS) & 455 (16.61\%) & 246 (8.98\%) & 209 (7.63\%) \\
\midrule
\textbf{Total Instances} & \textbf{2,740 (100.00\%)} & \textbf{1415 (51.64\%)} & \textbf{1325 (48.36\%)} \\
\bottomrule
\end{tabular}
\label{tab:owasp_dataset_description}
\end{table}

\noindent\textbf{Benchmark Dataset.}
To establish ground truth, we utilized the inherent labels provided by the OWASP Benchmark suite. 
We compared the aggregated SAST alerts against the benchmark's scorecard to categorize each alert as a True Positive (TP) or False Positive (FP).
The OWASP Benchmark Project is a widely recognized Java test suite designed to verify the speed and accuracy of vulnerability detection tools \cite{owaspbenchmark}, and has been used in numerous industries and for quantitative evaluation of SAST tools \cite{higuera2020owasp}. 
Prior studies on evaluating performance for static analysis, specifically on vulnerability detection, have widely used the OWASP Benchmark \cite{li2023comparison, khare2025understanding, akuthota2023vulnerability}.
Each test instance within the benchmark is a simple Java EE servlet, while the vulnerabilities within the benchmark are implemented and injected into programs manually \cite{li2023comparison}. 
Each test instance is mapped to specific Common Weakness Enumeration (CWE) Weaknesses, making it easier to automate and evaluate against SAST tool outputs. 
Furthermore, the evaluation scripts for most SAST tools are provided in the source \cite{owaspbenchmarksource}, thereby eliminating the possibility of tool misconfiguration. 
At the time of writing, the latest version of this benchmark is version 1.2. 
We use this version of the benchmark in our experiment. 
Table~\ref{tab:owasp_dataset_description} presents the benchmark statistics.

\noindent\textbf{Initial SAST Scanning.}
We selected four open-source SAST tools to generate an initial set of findings: CodeQL CLI v2.23.7~\cite{codeqlcli-v2237}, Semgrep v1.145.0~\cite{semgrep-v11450}, SonarQube Community v9.9.8~\cite{sonarqube-v998}, and Joern v4.0.454~\cite{joern-v40454}. 
These tools represent different analysis paradigms, including semantic query engines and syntactic pattern matchers. 
This selection also aligns with previous works \cite{li2023comparison,li2024evaluating}.

We scanned the benchmark with each tool using default rule sets to gather its alerts.
Each file contains only one vulnerable (or no) function within the benchmark framework.
If a tool generates an alert on a file and the scope of the finding is within the framework code, we consider the file as a positive, flagged by the tool. 

We aggregate the alerts by instance using a union-based approach: an instance is included in the candidate FP set if it is flagged as positive by at least one SAST tool.
This aggregation reduces computation costs while maintaining a high recall of potential vulnerabilities for agent verification.

Each tool adopts a slightly different running mechanism, specifically (1) \textbf{CodeQL.} We consider each "result" reported by \texttt{codeql database analyze} ran after \texttt{codeql database create} with default rules as a positive finding. (2) \textbf{Semgrep.} We consider each "result" reported by \texttt{semgrep scan} with default rules as a positive finding. (3) \textbf{SonarQube.} We consider each "issue" under "VULNERABILITY" type regardless of its severity, as a positive finding. We do not consider "security hotspot" as a positive finding. (4) \textbf{Joern.} We consider each line of "Result" reported by \texttt{joern-scan} as a positive finding.

\noindent\textbf{Agentic Frameworks.}
We then provided these aggregated findings to three LLM-based agent frameworks:
\textsc{Aider} v0.86.1~\cite{aider,aider-v0861},
\textsc{OpenHands} Agent SDK v1.5.2~\cite{wang2025openhands,openhands-sdk-v152},
and \textsc{SWE-agent} commit 1d3cfb~\cite{yang2024sweagent,sweagent-commit-1d3cfb}.
They represent three interaction styles: pair-programming assistance (\textsc{Aider}), a developer-like workspace with editor/terminal/browser (\textsc{OpenHands}), and repository-level autonomous repair (\textsc{SWE-agent}).
We utilized three state-of-the-art models as the backbone for these systems: Claude Sonnet 4 (\texttt{claude-sonnet-4-20250514}), DeepSeek Chat (\texttt{deepseek-chat}), and GPT-5 (\texttt{gpt-5-2025-08-07}).
Each agent was provided with: (1) the specific warning messages generated by the SAST tools for the target file, (2) access to the test code, and (3) a prompt instructing the system to "verify if this static analysis finding represents a real vulnerability or an FP". 
To ensure fair comparison, we disable browser use and web browsing tools where possible for each agent. 
\mr{R6}{We made this design choice for two reasons.
First, it keeps the evaluation focused on code reasoning rather than on whether an agent can find useful information on the web.
Second, it protects benchmark integrity: unrestricted web access could allow an agent to retrieve benchmark scorecards, issue discussions, fix commits, or other external artifacts that directly reveal the expected ground truth.
Such retrieval would change the task from code-based SAST triage to web-based answer lookup.
To verify that this constraint did not materially bias our results, we conducted a paired ablation experiment using \textsc{OpenHands} with Claude Sonnet 4. We constructed the 50-case subset from the completed browser-disabled OWASP run: using a fixed Python random seed, we shuffled the completed case list, selected the first 50 cases, and re-ran exactly those instances with browser access enabled.
Browser-disabled achieved an 86.0\% FP identification rate (43/50), while browser-enabled achieved 82.0\% (41/50), indicating that browser access provided no measurable advantage.
}
We report metrics for each combination of model and agent.

\noindent\textbf{Vanilla LLM.} 
We employ zero-shot prompting on the LLMs to serve as a baseline. 
The prompt establishes a security-oriented reviewer persona and provides the model with necessary context, including the repository structure, the complete source code of the flagged file, and the raw findings from the SAST tool. 
To maintain consistency with the agentic setups, the prompt specifies a read-only environment and requires a triage decision based on the provided inputs. 
The complete prompt template and its variants are available in our replication package.

\noindent\textbf{Metrics.} 
We evaluate model performance by comparing the predicted labels against the benchmark ground truth. Results are categorized into True Positives (TP), False Positives (FP), True Negatives (TN), and False Negatives (FN) based on the benchmark ground truth. 
Specifically, a TP represents a correctly identified vulnerability, while an FP occurs when the tool reports an alert on an instance labeled as negative in the ground truth. Conversely, a TN indicates the correct identification of a non-vulnerable instance, and an FN occurs when the tool fails to detect a known vulnerability.

We also define False Positive Rate (FPR).
FPR presents the percentage of non-vulnerable instances that were incorrectly flagged as positive by at least one SAST tool. 
    
    

\subsection{Methodology for RQ2}
\label{rq2:methodology}

\begin{figure}[htbp]
  \centering
  \begin{minipage}{0.48\textwidth}
    \centering
    \includegraphics[width=\linewidth]{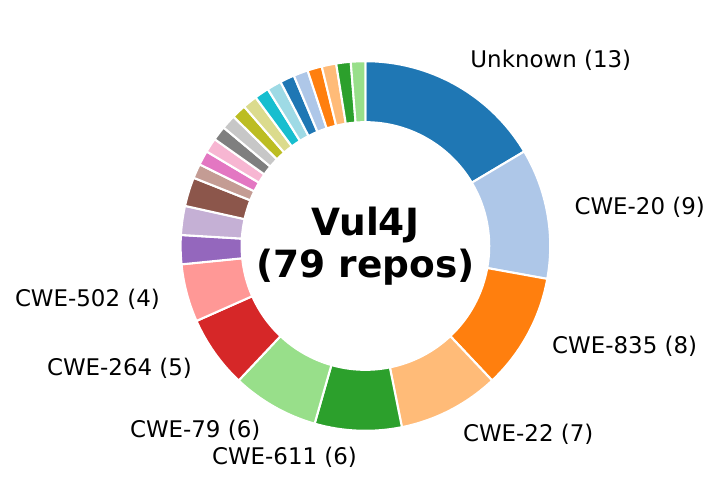}
    \caption{Distribution of CWE identified in the Vul4J dataset.}
    \label{fig:vul4j_cwe_distribution}
  \end{minipage}
  \hfill
  \begin{minipage}{0.48\textwidth}
    \centering
    \includegraphics[width=\linewidth]{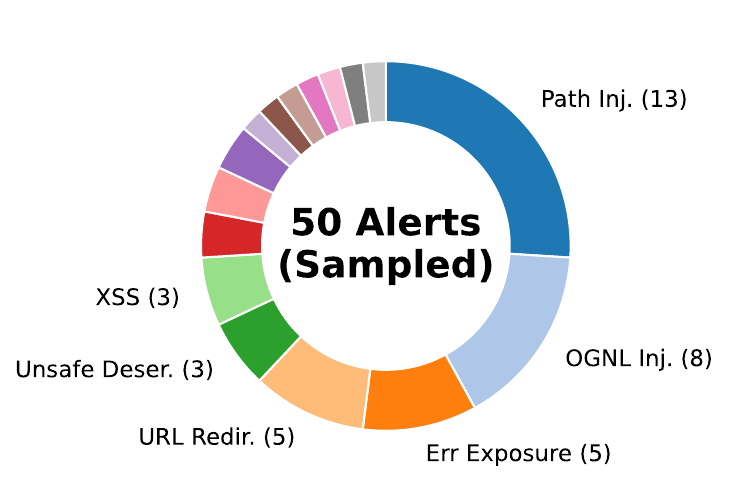}
    \caption{Distribution of CodeQL rule IDs within the Vul4J sample used for RQ2· ($n=50$).}
    \label{fig:vul4j_rule_distribution}
  \end{minipage}
\end{figure}

\noindent\textbf{Benchmark Dataset.} We evaluate RQ2 using the Vul4J dataset \cite{bui2022vul4j}, which consists of 79 reproducible Java vulnerabilities derived from Project KB \cite{ponta2019projectkb}.
Those vulnerabilities spanned across 51 distinct repositories, spanning a diverse set of real-world Java projects, covering enterprise web frameworks and middleware (e.g., Spring, Struts, CXF), security libraries (e.g., Shiro, ESAPI), data parsing/serialization components (e.g., fastjson, Jackson, XStream), and DevOps infrastructure such as Jenkins and its plugin ecosystem. 
The vulnerabilities cover 25 CWEs in total. A breakdown of the CWE categories is illustrated in Figure~\ref{fig:vul4j_cwe_distribution}.
To notice, the category of \textit{Not Mapping} refers to when the authors of the Vul4J dataset failed to retrieve the CWE information for the vulnerabilities.

The dataset is provided as 79 branches in the original dataset hosted on GitHub. 
Each branch represents a specific vulnerable version of a certain repository in question, which is a data point in the dataset.  
For evaluation, we checked out each of them and manually copied them as 79 distinct folders. 
These 79 branches have 6,786,406 source lines of code in total, where the mean and median for each branch are 85,903 and 42,136, respectively. 
We scanned these 79 cases using CodeQL, resulting in a total of 3,426 alerts, where the mean and median are 43 and 16.



\noindent\textbf{Manual Triage.} Due to the large volume of the alerts, we sampled 50 alerts from these for manual triage.
We performed the sampling using Python's \textit{random} library with a fixed seed to ensure reproducibility.
This sampled set represents 26 projects and covers 16 unique CodeQL rules, accounting for 37\% of the total rules observed in the original scan. A breakdown of the CodeQL rules observed in these samples is illustrated in Figure~\ref{fig:vul4j_rule_distribution}.

To establish ground truth labels for these alerts, we followed a two-step process. 
First, we automatically matched CodeQL alerts against the human-written patches provided in Vul4J. 
Following previous work \cite{li2023comparison}, we employed method-level matching: any alert falling within a patched method with a matching CWE type was automatically marked as a TP. 
In our sampled 50 alerts, 2 of them were automatically labeled as TP with this strategy.
For the remaining 48 unlabeled alerts, the first author served as the annotator to manually inspect each sample. 
The annotator has 6 years of experience in the security field and programming.
The annotation process can be described as follows:
The annotator iterates through sampled alerts.
For each alert, the annotator:
(1) Browse the project associates with this alert and locate the file;
(2) Study the actual CodeQL rule and description and code flow to have an understanding of what this alert is about;
(3) Perform analysis including but not limited to: data flow analysis, checking for known insecure patterns, searching for issues mentioning this file, refer to other static analyzers;
(4) Consult to other experts if the annotator does not have confidence in the nature of this alert after the annotator perform the above steps. Otherwise, continue;
(5) Label it as a true vulnerability, i.e., TP, if it accurately represented its description and category; otherwise, we labeled it as an FP. The whole process took about 20 man-hours. Finally, among the 50 alerts, 31 are labeled as FPs, and 19 are labeled as TPs, yielding an initial FP identification precision of 62\%.

We then provided these labeled alerts to three agent frameworks:  \textsc{Aider}, \textsc{OpenHands}, and \textsc{SWE-agent}.
Similarly, these frameworks utilized Claude Sonnet 4, DeepSeek Chat, and GPT-5, as their backbone models. 
Each agent received: (1) the CodeQL warning message, (2) access to the full repository context for cross-file reasoning, and (3) a prompt instructing the agent to "verify if this static analysis finding represents a real vulnerability or an FP", following the procedure in RQ1~\ref{rq1:methodology}. 
Finally, we compare the results produced by these agents to their backbone model, i.e., the vanilla LLM, with zero-shot prompting, the same as in RQ1 (Section~\ref{rq1:methodology}). 
We conduct qualitative analysis in Section~\ref{sec:discussion}.

\noindent\textbf{Metrics.}
Unlike RQ1, which reports residual FPR from a tool-output perspective, RQ2 evaluates alert-level triage as a binary classification task.
We treat SAST FPs as the positive class because the operational goal is to identify alerts that can be safely filtered.
To maintain terminology consistency and address the safety implications of agentic filtering, we define the outcome categories as follows: 
(1) Identified FP: The agent successfully identifies an SAST FP.
(2) Missed FP: The agent fails to identify an SAST FP.
(3) Mis-identified FP: A safety violation where the agent incorrectly classifies a true vulnerability as an FP.
(4) Correct Retention: The agent successfully identifies and preserves a true vulnerability.
Under this framework, recall is the FP identification rate, precision is the probability that an alert classified as an FP is indeed an FP, and F1 summarizes the precision--recall trade-off.

\mr{R1 (1)}{
\subsection{Methodology for RQ3}
\label{rq3:methodology}
Unlike RQ2's full Java comparison, RQ3 is a focused mechanism and generalization study.
We use the strongest RQ1 configuration, Claude Sonnet 4 with \textsc{SWE-agent}, and compare it with targeted ablations and non-agentic baselines on OSS-Fuzz.
This tests whether agentic gains persist in a contamination-free C/C++ setting and which capabilities explain them.

\noindent\textbf{C/C++ Evaluation.}
To address concerns about possible data leakage and to extend our evaluation beyond Java, we built a post-cutoff C/C++ alert pool from Google's OSS-Fuzz~\cite{ossfuzz} continuous fuzzing service.
Starting from the most recent records in the public OSV export for OSS-Fuzz, we scanned backward through OSS-Fuzz records until we identified recent C/C++ projects whose fixing commits postdated the known training cutoff of Claude Sonnet 4 by 10--12 months and whose vulnerable snapshots could be reconstructed.
For each retained OSV record, we used the OSS-Fuzz/OSV metadata to obtain the fixing commit, derived the corresponding vulnerable commit as the single parent of that fixing commit, and reconstructed the corresponding vulnerable repository snapshot.
The final sampled alerts cover five projects (vulnerable fixing commit time in bracket): mpv (January 26, 2026), libhevc (February 14, 2026), libical (February 21, 2026), gpsd (March 11, 2026), and quickjs (March 21--23, 2026).
We scanned the vulnerable versions of these projects using CodeQL with default C/C++ rules, generating a pool of 804 eligible alerts after filtering out test/generated files and non-security/non-CWE-tagged findings.
From this pool, we sampled 50 alerts using a fixed random seed (seed=42), ensuring reproducibility.
The sample covers all five projects with the following distribution: mpv (11 alerts), gpsd (16 alerts), libical (9 alerts), libhevc (13 alerts), and quickjs (1 alert). 
The sampled alerts cover 11 CodeQL C/C++ rule IDs. According to CodeQL's full CWE coverage documentation~\cite{codeql_full_cwe_coverage}, these rules map to 25 security-relevant CWE IDs. These CWE IDs span high-level weakness types including path traversal and path manipulation (CWE-22/23/36/73/610/642/706), numeric overflow/conversion/calculation (CWE-190/197/681/682/704), access-control and permission weaknesses (CWE-284/285/668/693/732), initialization and nullness issues (CWE-457/476/665), suspicious comments (CWE-546), unused/dead code (CWE-561/563), and dangerous function use (CWE-676).
To establish ground truth, two independent raters, each with 6+ years of programming experience, labeled all 50 alerts.
The raters achieved 86.0\% raw agreement (43/50 cases), with Cohen's $\kappa = 0.725$, indicating substantial agreement.
The 7 disagreements were resolved through discussion, yielding a final ground truth of 22 FPs and 28 TPs.

\noindent\textbf{Experimental Configurations.}
On this OSS-Fuzz dataset, we evaluated three types of configurations to isolate the sources of agentic gains:~\noindent\textbf{(1) Full agent configurations:} We evaluate the most capable configuration in RQ1 (\textsc{SWE-agent} with Claude Sonnet 4).~\noindent\textbf{(2) Ablation variants:} We systematically removed four capabilities from the full configuration: multi-turn interaction, cross-file navigation, tool-based verification (e.g., arithmetic checks), and configuration-file access.
These dimensions were motivated by prior work where ReAct emphasizes interleaving reasoning and action, while \textsc{SWE-agent} emphasizes repository navigation and tool-enabled interaction and by our own trajectory analysis, which identified configuration validation as a recurring success pattern.~\noindent\textbf{(3) Stronger baselines:} We evaluated three additional baselines beyond vanilla prompting: (a) LLM4SA-style~\cite{wen2024llm4sa} prompting, where we reuse the LLM4SA task instruction and output-label format, fill its \textit{Bug Report} field with the CodeQL alert JSON, and fill its \textit{Code Snippet} field with the complete reported source file, but do not reproduce LLM4SA's separate program-dependence context-extraction pipeline; (b) context-augmented vanilla LLM, which provides cross-file context via heuristic-based searching including call-graph neighbors, direct includes, CodeQL generated code-flow and related locations; and (c) oracle-context vanilla LLM, which receives the exact same files that our full configuration accessed during its successful runs.
The oracle-context baseline allows us to isolate the contribution of iterative reasoning from context volume.
}

\noindent\textbf{Metrics.}
RQ3 uses the same alert-level FP-filtering metrics as RQ2, with SAST FPs as the positive class.
We report accuracy, precision, recall, and F1; recall corresponds to the FP identification rate.



%







\subsection{Methodology for RQ4}
To understand the performance gap between agentic frameworks and vanilla prompting, we conduct a manual inspection of execution trajectories. Our analysis focuses on two distinct sets of cases:

\noindent\textbf{Success Analysis:} We examine 256 trajectories where the standalone LLM incorrectly predicts a TP but \textsc{SWE-agent} (with Claude Sonnet 4) correctly identifies a FP. These cases cover 11 OWASP categories, primarily CWE-330 (126 cases), CWE-78 (35 cases), and CWE-79 (20 cases).

\noindent\textbf{Failure Analysis:} We analyze 103 failed trajectories where the agent misses an FP. This set includes 59 cases where both \textsc{OpenHands} and \textsc{SWE-agent} failed, plus targeted inspection of the dominant failure categories: CWE-327 (47 cases) and CWE-614 (10 cases).
For each trajectory, we record the number of steps, the files accessed, and the specific tools utilized. 
We then categorize the underlying reasoning patterns into success patterns (P1--P4) and failure modes (F1--F3) based on the agent's interaction with the codebase and configuration files.








\section{Results}
\label{sec:results}



\subsection{RQ1: Filtering Performance of LLM-Based Agents} 
\label{rq1:result}

\subsubsection{FPs Produced by SAST Tools.}
\begin{table*}[!th]
\centering

\begin{minipage}[t]{0.5\textwidth}\vspace{0pt}
  \centering
  \captionsetup{type=table,skip=2pt}
  \caption{Initial Performance of SAST Tools Across All Instances. The Union represents the candidate pool provided to LLM agents for filtering.}
  \label{tab:sast_initial_scan_performance}
  \begin{adjustbox}{max width=\linewidth}
    \begin{tabular}{lrrrrrr}
        \toprule
        Tool & Accuracy & Precision & Recall & F1 & FPR \\
        \midrule
        CodeQL & \textbf{65.5\%} & \textbf{60.3\%} & \textbf{97.0\%} & \textbf{74.4\%} & \underline{68.2\%} \\
        \midrule
        Joern & 49.1\% & 54.7\% & 8.2\% & 14.3\% & \textbf{7.2}\% \\
        \midrule
        Semgrep & \underline{58.9\%} & \underline{56.3\%} & 90.4\% & \underline{69.4\%} & 74.8\% \\
        \midrule
        SonarQube & 52.0\% & 51.9\% & \underline{95.6\%} & 67.3\% & 94.6\% \\
        \bottomrule
        \end{tabular}
  \end{adjustbox}
\end{minipage}\hfill
\begin{minipage}[t]{0.5\textwidth}\vspace{0pt}
  \centering
  \includegraphics[width=0.92\linewidth]{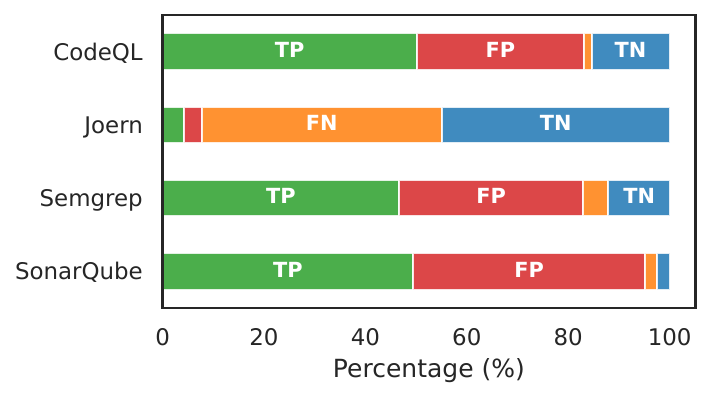}
  \captionsetup{type=figure,skip=2pt}
  \caption{Distribution of SAST alerts (Overall Scope).}
  \label{fig:sast_findings_distribution}
\end{minipage}

\end{table*}

Table~\ref{tab:sast_initial_scan_performance} summarizes initial SAST performance; bold and underline mark the best and second-best results.
Figure~\ref{fig:sast_findings_distribution} shows each tool's alert distribution.
CodeQL achieves the highest F1 (74.4\%) and precision (60.3\%), but still produces 904 FPs, covering 33.0\% of benchmark cases and 68.2\% of ground-truth non-vulnerable cases.
Semgrep and SonarQube have lower precision and recall than CodeQL, with recall rates of 90.4\% and 95.6\%.
SonarQube produces the most FPs (1,254; 45.8\% of the benchmark) and flags 94.6\% of non-vulnerable cases as positive.
Joern is conservative, producing only 96 FPs but missing most vulnerabilities, with 8.2\% recall.
This may stem from Joern's limited default Java query set, which contains 6 rule files and 7 vulnerable-code detection functions~\cite{joernscandocs,joernscanjavarulesets}.


\begin{figure*}[h]
\centering
    \begin{subfigure}[b]{0.48\textwidth}
        \centering
        \includegraphics[width=\textwidth]{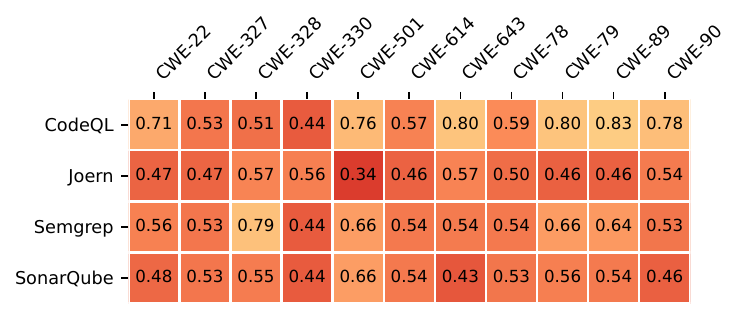}
        \caption{Accuracy across different CWE categories.}
\label{fig:accuracy_heatmap_sast_tool_performance_aggerated_by_cwe}
    \end{subfigure}
    \hfill
    \begin{subfigure}[b]{0.48\textwidth}
        \centering
        \includegraphics[width=\textwidth]{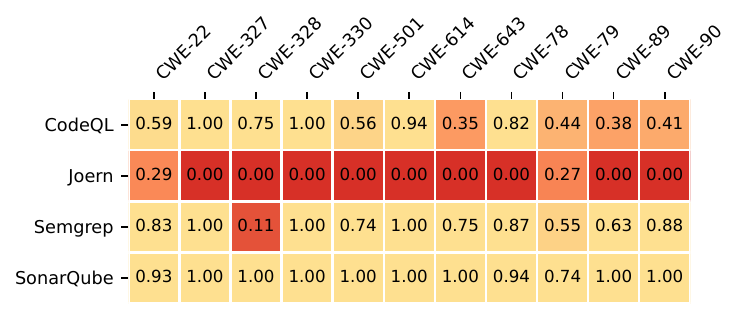}
        \caption{FPR across different CWE categories.}
        \label{fig:fpr_heatmap_sast_tool_performance_aggerated_by_cwe}
    \end{subfigure}
    \caption{Performance comparison of four SAST tools across various CWEs. (a) Heatmap illustrating accuracy. (b) Heatmap illustrating FPR, where values of 1.00 indicate that tools flag all non-vulnerable instances in those categories.}
    \label{fig:overall_performance_heatmap}
\end{figure*}

Figure~\ref{fig:accuracy_heatmap_sast_tool_performance_aggerated_by_cwe} shows that the detection capabilities vary across tools and vulnerability types.
CodeQL consistently demonstrates robust detection performance, achieving the highest accuracy in the majority of categories, specifically peaking at 0.83 for CWE-89 (SQL Injection) and 0.80 for CWE-78 (OS Command Injection) and CWE-643 (XPATH Injection). 
Semgrep shows competitive performance in specific categories, notably outperforming other tools in CWE-328 (Weak Hashing) with an accuracy of 0.79.
Conversely, Joern and SonarQube exhibit moderate to lower accuracy across the board, with Joern dropping as low as 0.34 for CWE-501 (Trust Boundary Violation).

Figure~\ref{fig:fpr_heatmap_sast_tool_performance_aggerated_by_cwe} highlights a high volume of FPs generated by traditional SAST tools. 
In our initial scan, SonarQube and Semgrep exhibit excessively high FPRs. 
SonarQube, in particular, reaches an FPR of 1.00 across nearly all evaluated CWEs, indicating that for these categories, the tool flags almost every non-vulnerable test instances as a vulnerability. 
Similarly, Semgrep shows an FPR of 1.00 for multiple categories, including CWE-327 and CWE-330. 
While CodeQL achieves the highest accuracy, FPR reveals that this comes at the cost of precision. 
Its FPR reaches 1.00 for CWE-327 and CWE-330, and remains above 0.80 for injection flaws like CWE-78 (0.82). 
In comparison, Joern maintains an FPR of 0.00 for the majority of CWEs. 
While this can indicate high precision, its lower accuracy may suggest a conservative analysis strategy that likely suffers from a high false negative rate, missing actual vulnerabilities to avoid false alarms.

\begin{table*}[!th]
\begin{minipage}{0.58\textwidth}
\centering
\small
\caption{Overall FP reduction performance on the OWASP benchmark (FPR, lower is better).}
\begin{adjustbox}{max width=\textwidth}
\begin{tabular}{llr}
    \toprule
    Model & Agent & FPR (compared to SAST) \\
    \midrule
    \multirow{4}{*}{Claude Sonnet 4} 
        & \textsc{Aider} & 14.3\% (\textcolor{red}{$\downarrow$} 84.1\%) \\
        & \textsc{OpenHands} & 14.9\% (\textcolor{red}{$\downarrow$} 83.5\%) \\
        & \textsc{SWE-agent} & 6.3\% (\textcolor{red}{$\downarrow$} 92.1\%) \\
        & Vanilla LLM & 23.0\% (\textcolor{red}{$\downarrow$} 75.3\%) \\
    \midrule
    \multirow{4}{*}{DeepSeek Chat} 
        & \textsc{Aider} & 13.2\% (\textcolor{red}{$\downarrow$} 85.1\%) \\
        & \textsc{OpenHands} & 15.8\% (\textcolor{red}{$\downarrow$} 82.6\%) \\
        & \textsc{SWE-agent} & 13.1\% (\textcolor{red}{$\downarrow$} 85.2\%) \\
        & Vanilla LLM & 11.2\% (\textcolor{red}{$\downarrow$} 87.1\%) \\
    \midrule
    \multirow{4}{*}{GPT-5} 
        & \textsc{Aider} & 20.3\% (\textcolor{red}{$\downarrow$} 78.0\%) \\
        & \textsc{OpenHands} & 16.3\% (\textcolor{red}{$\downarrow$} 82.0\%) \\
        & \textsc{SWE-agent} & 14.1\% (\textcolor{red}{$\downarrow$} 84.2\%) \\
        & Vanilla LLM & 20.4\% (\textcolor{red}{$\downarrow$} 78.0\%) \\
    \bottomrule
\end{tabular}
\end{adjustbox}
\label{tab:r1_fp_reduction_performance}
\end{minipage}
\hfill
\begin{minipage}{0.38\textwidth}
\centering
\vspace{-6pt}
\includegraphics[width=0.95\textwidth]{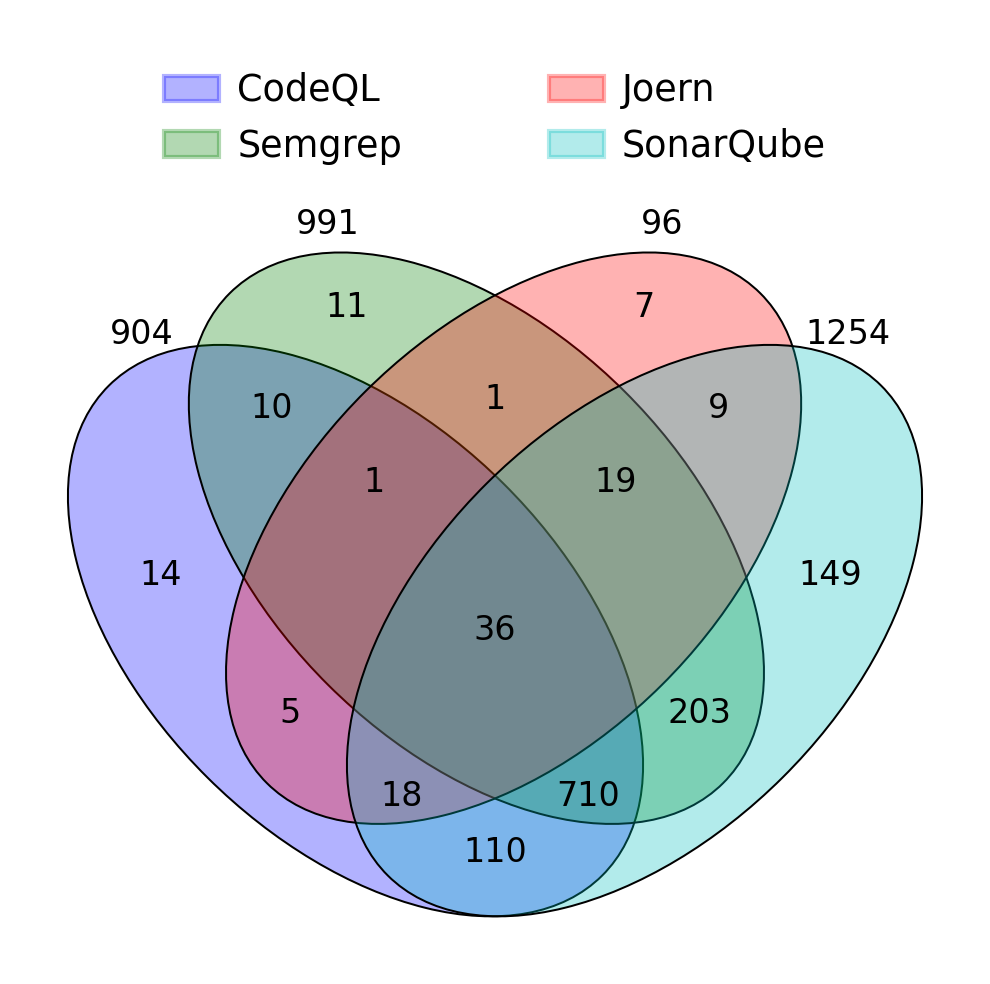}
\vspace{-6pt}
\captionof{figure}{Distinct FP cases detected by each tool.}
\label{fig:distinct_false_positive_files_detected_by_each_tool}
\end{minipage}
\vspace{-2mm}
\end{table*}

To understand whether this noise is tool-specific or correlated, we analyzed the overlap of FP instances across scanners. 
As shown in Figure~\ref{fig:distinct_false_positive_files_detected_by_each_tool}, FPs are largely shared rather than unique.
Only a small fraction of noise is exclusive to a single tool, e.g., 14 instances unique to CodeQL and 11 unique to Semgrep,  while SonarQube acts as a near-superset of the FPs reported by other scanners.
More importantly, shared-error regions remain large: 36 FP instances are flagged by all four tools, and CodeQL, Semgrep, and SonarQube jointly flag a dominant block of 710 instances.

\subsubsection{Effectiveness of Agentic FP Filtering}
We evaluate LLM agents on the union of SAST alerts, where at least one tool flags $1,303$ of $1,325$ non-vulnerable instances, yielding a 98.3\% baseline FP rate.
Table~\ref{tab:r1_fp_reduction_performance} reports the remaining FPR after agent verification.
Among all configurations, \textsc{SWE-agent} paired with Claude Sonnet 4 achieves the highest filtering performance. 
This setup reduces the remaining FPR to $6.3\%$, representing a $92.1\%$ reduction from the original rate. 
This performance exceeds that of \textsc{OpenHands} at $14.9\%$ and \textsc{Aider} at $14.3\%$. 
When using GPT-5, \textsc{SWE-agent} remains the top-performing agent with a $14.1\%$ FPR, though the performance gap among frameworks decreases. 
For DeepSeek Chat, agent performance ranges between $13.1\%$ and $15.8\%$, suggesting that increased complexity in agentic loops does not consistently result in improved filtering.
Overall gains are strongly backbone-dependent.
For Claude Sonnet 4, \textsc{SWE-agent} reduces the remaining FPR from $23.0\%$ with vanilla prompting to $6.3\%$, demonstrating a clear advantage for iterative logic. 
For GPT-5, \textsc{SWE-agent} also improves upon the vanilla baseline, reducing the FPR from $20.4\%$ to $14.1\%$. 
In contrast, DeepSeek Chat shows no consistent improvement from an agentic workflow; the vanilla baseline already achieves a low remaining FPR of $11.2\%$, while its agentic counterparts yield comparable or higher residual error rates.
\subsubsection{Categorical Analysis across CWEs}
\label{sec:categorical-analysis-across-cwes}

\begin{figure*}[h]
  \centering
  \includegraphics[width=\linewidth]{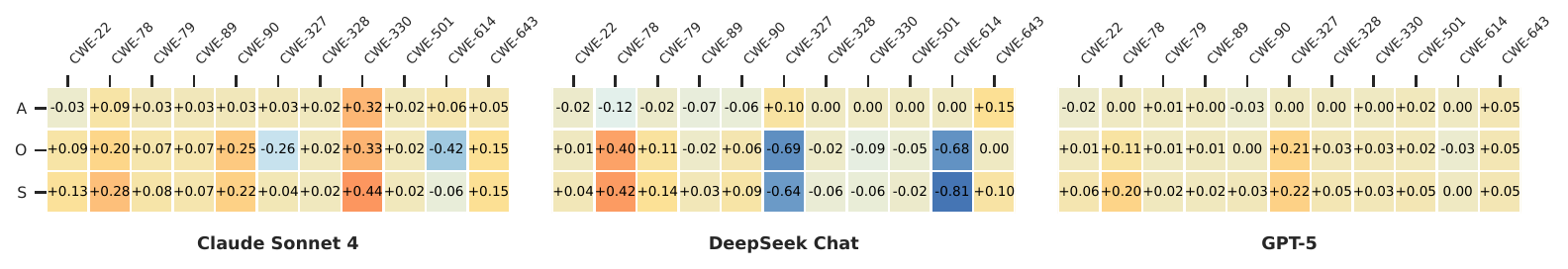}
  \caption{CWE-level change in FP filtering success over vanilla prompting. Values are percentage-point changes; red indicates gains and blue for regressions. (A=\textsc{Aider}, O=\textsc{OpenHands}, S=\textsc{SWE-agent}.)}
  \label{fig:agentic_gain_heatmap}
\end{figure*}

A CWE-level analysis indicates that agentic workflows do not provide uniform benefits across different vulnerability types. 
Figure~\ref{fig:agentic_gain_heatmap} illustrates the performance difference between the three agents and the vanilla LLM baseline, measured in percentage points (pp) of FP filtering success. 

Under Claude Sonnet 4, all three frameworks improve filtering for high-volume injection categories and CWE-330, though the degree of improvement varies. \textsc{SWE-agent} is the most consistently beneficial, improving performance in 10 out of 11 CWEs. Its largest improvements occur in CWE-330, with a $44.4$,pp increase from $51.6\%$ to $96.0\%$, and CWE-78, which increases by $28.5$,pp from $69.9\%$ to $98.4\%$. \textsc{OpenHands} also shows improvements in CWE-78 and CWE-330, yet it exhibits a regression in CWE-614, where performance drops by $41.9$,pp from $74.2\%$ to $32.3\%$, and in CWE-327, which decreases by $26.0$,pp from $55.2\%$ to $29.2\%$. \textsc{Aider} remains more stable on policy and cryptographic categories; for instance, its CWE-614 performance improves to $80.6\%$, although its gains on high-noise blocks are smaller than those of \textsc{SWE-agent}.

For DeepSeek Chat, the effects of an agentic workflow are complex and polarized. \textsc{OpenHands} and \textsc{SWE-agent} improve injection-heavy categories, particularly CWE-78, with gains of $39.9$,pp and $42.3$,pp respectively. However, these same agents demonstrate a sharp degradation in policy and cryptographic categories. Performance in CWE-614 drops by $67.7$,pp and $80.6$,pp, while CWE-327 success decreases by $69.0$,pp and $63.8$,pp. Conversely, \textsc{Aider} preserves perfect performance in CWE-614 at $100\%$ and increases CWE-327 success to $100\%$, but suffers a $12.2$,pp regression in CWE-78.

GPT-5 exhibits almost no regressions when an agentic workflow is applied. Both \textsc{OpenHands} and \textsc{SWE-agent} raise CWE-327 filtering success from $1.7\%$ to over $22\%$, and improve CWE-78 performance by $10.5$,pp and $20.3$,pp respectively. For categories with high baseline success, such as CWE-79, CWE-89, CWE-614, and CWE-643, all frameworks operate near the performance ceiling, resulting in smaller marginal gains.

\mynote{\textbf{Answer to RQ1.} LLM-based agents reduce the OWASP residual FPR from 98.3\% to as low as 6.3\%, but the gain depends on backbone model and CWE. Claude Sonnet 4 and GPT-5 benefit from agentic behaviour, while DeepSeek Chat shows no net gain; residual FPs concentrate in policy-based categories like weak cryptography.}

\subsection{RQ2: Performance of LLM Agents in Real-World \mrm{Java} Scenarios}
\label{rq2:result}

\begin{table}[h]
\centering
\small
\caption{Performance in selected Vul4J projects (\%). Identification metrics refer to the agent's ability to classify SAST FPs correctly while avoiding misidentification of true vulnerabilities.}
\label{tab:real_world_performance}
\begin{tabular}{llcccccccc}
\toprule
Model & Agent & \makecell{Iden. \\ FP} & \makecell{Corr. \\ Ret.} & \makecell{Mis-id. \\ FP} & \makecell{Missed \\ FP} & Acc & Prec & Rec & F1 \\
\midrule
\multirow{4}{*}{Claude Sonnet 4} & \textsc{Aider} & 46.9 & 30.6 & 8.2 & 14.3 & \textbf{77.6} & 85.2 & 76.7 & \underline{80.7} \\
 & \cellcolor[HTML]{e1dcef}\textsc{OpenHands} & \cellcolor[HTML]{e1dcef}42.0 & \cellcolor[HTML]{e1dcef}34.0 & \cellcolor[HTML]{e1dcef}4.0 & \cellcolor[HTML]{e1dcef}20.0 & \cellcolor[HTML]{e1dcef}\underline{76.0} & \cellcolor[HTML]{e1dcef}\underline{91.3} & \cellcolor[HTML]{e1dcef}67.7 & \cellcolor[HTML]{e1dcef}77.8 \\
 & \textsc{SWE-agent} & 38.3 & 29.8 & 10.6 & 21.3 & 68.1 & 78.3 & 64.3 & 70.6 \\
 & \cellcolor[HTML]{e1dcef}Vanilla LLM & \cellcolor[HTML]{e1dcef}38.0 & \cellcolor[HTML]{e1dcef}28.0 & \cellcolor[HTML]{e1dcef}10.0 & \cellcolor[HTML]{e1dcef}24.0 & \cellcolor[HTML]{e1dcef}66.0 & \cellcolor[HTML]{e1dcef}79.2 & \cellcolor[HTML]{e1dcef}61.3 & \cellcolor[HTML]{e1dcef}69.1 \\
\midrule
\multirow{4}{*}{DeepSeek Chat} & \textsc{Aider} & 34.0 & 34.0 & 4.0 & 28.0 & 68.0 & 89.5 & 54.8 & 68.0 \\
 & \cellcolor[HTML]{e1dcef}\textsc{OpenHands} & \cellcolor[HTML]{e1dcef}24.0 & \cellcolor[HTML]{e1dcef}38.0 & \cellcolor[HTML]{e1dcef}0.0 & \cellcolor[HTML]{e1dcef}38.0 & \cellcolor[HTML]{e1dcef}62.0 & \cellcolor[HTML]{e1dcef}\textbf{100.0} & \cellcolor[HTML]{e1dcef}38.7 & \cellcolor[HTML]{e1dcef}55.8 \\
 & \textsc{SWE-agent} & 31.2 & 35.4 & 4.2 & 29.2 & 66.7 & 88.2 & 51.7 & 65.2 \\
 & \cellcolor[HTML]{e1dcef}Vanilla LLM & \cellcolor[HTML]{e1dcef}42.0 & \cellcolor[HTML]{e1dcef}16.0 & \cellcolor[HTML]{e1dcef}22.0 & \cellcolor[HTML]{e1dcef}20.0 & \cellcolor[HTML]{e1dcef}58.0 & \cellcolor[HTML]{e1dcef}65.6 & \cellcolor[HTML]{e1dcef}67.7 & \cellcolor[HTML]{e1dcef}66.7 \\
\midrule
\multirow{4}{*}{GPT-5} & \textsc{Aider} & 61.0 & 14.6 & 19.5 & 4.9 & 75.6 & 75.8 & \underline{92.6} & \textbf{83.3} \\
 & \cellcolor[HTML]{e1dcef}\textsc{OpenHands} & \cellcolor[HTML]{e1dcef}58.3 & \cellcolor[HTML]{e1dcef}12.5 & \cellcolor[HTML]{e1dcef}25.0 & \cellcolor[HTML]{e1dcef}4.2 & \cellcolor[HTML]{e1dcef}70.8 & \cellcolor[HTML]{e1dcef}70.0 & \cellcolor[HTML]{e1dcef}\textbf{93.3} & \cellcolor[HTML]{e1dcef}80.0 \\
 & \textsc{SWE-agent} & 53.2 & 21.3 & 19.1 & 6.4 & 74.5 & 73.5 & 89.3 & 80.6 \\
 & \cellcolor[HTML]{e1dcef}Vanilla LLM & \cellcolor[HTML]{e1dcef}57.1 & \cellcolor[HTML]{e1dcef}14.3 & \cellcolor[HTML]{e1dcef}19.0 & \cellcolor[HTML]{e1dcef}9.5 & \cellcolor[HTML]{e1dcef}71.4 & \cellcolor[HTML]{e1dcef}75.0 & \cellcolor[HTML]{e1dcef}85.7 & \cellcolor[HTML]{e1dcef}80.0 \\
\bottomrule
\end{tabular}
\end{table}

Table~\ref{tab:real_world_performance} shows that FP mitigation performance remains highly backbone-dependent. 
For Claude Sonnet 4, agentic workflows yield a clear advantage over vanilla prompting. 
\textsc{Aider} achieves the most balanced performance with a recall of $76.7\%$ and precision of $85.2\%$. 
\textsc{OpenHands} adopts a more conservative strategy, yielding the highest precision at $91.3\%$ but a lower rate of identified FPs at $67.7\%$. 
These results indicate that for Claude, agentic workflows increase the efficacy of noise reduction, though different framework designs occupy different positions on the spectrum between safety and aggressiveness.

For DeepSeek Chat, the trade-off between noise removal and vulnerability retention is more pronounced. The vanilla LLM baseline achieves a recall of $67.7\%$ but maintains a low precision of $65.6\%$, indicating a higher frequency of mis-identified FPs. 
Conversely, \textsc{OpenHands} reaches $100\%$ precision but only identifies $38.7\%$ of the FPs, leaving noise unfiltered. 
\textsc{Aider} provides a balanced alternative with $89.5\%$ precision and $54.8\%$ recall. 
These outcomes highlight that a high rate of identified FPs is insufficient if it is achieved by incorrectly discarding actual security flaws.

For GPT-5, the recall is consistently high across all configurations and ranges from $85.7\%$ to $93.3\%$, which narrows the gap between agentic workflows and vanilla prompting. 
\textsc{Aider} yields the highest F1-score of $83.3\%$ with a $92.6\%$ recall and $75.8\%$ precision. 
\textsc{OpenHands} marginally increases the recall to $93.3\%$ but at a lower precision of $70.0\%$. 
This suggests that for highly capable backbone models, additional agentic scaffolding may yield diminishing returns in noise mitigation, and the primary differentiator shifts to how effectively the framework avoids mis-identifying true vulnerabilities.

\mynote{\textbf{Answer to RQ2.} On real-world Java CodeQL alerts, agents can identify up to 93.3\% of FPs, but the precision--recall trade-off remains backbone-dependent. Claude benefits most from agentic frameworks, GPT-5 shows smaller gains over vanilla prompting, and DeepSeek trades recall for precision.}

\mr{R1 (2), R2, R3}{
\subsection{RQ3: Agentic Capabilities and Post-Cutoff Generalization}
\label{rq3:result}
Table~\ref{tab:ossfuzz_results} presents the complete results for OSS-Fuzz, including the full \textsc{SWE-agent}, four ablation variants, and three additional baselines, all using Claude Sonnet 4 as the backbone model. On this contamination-free C/C++ dataset, the full \textsc{SWE-agent} identifies 95.5\% of FPs while maintaining 95.5\% precision, compared with a 36.4\% FP identification rate for vanilla prompting.

The ablation results reveal a capability hierarchy rather than a uniform benefit from all agentic behaviour.
\textbf{Cross-file navigation} is the most important capability: removing it drops accuracy from 96.0\% to 44.0\% and F1 from 95.5\% to 62.5\%.
Among the 26 full-correct cases missed by this ablation, all become \texttt{UNKNOWN}, showing that the agent can no longer collect enough evidence for a grounded verdict.
For example: in \texttt{mpv@eadba44}, the integer-overflow alert at \path{audio/filter/af_scaletempo2_internals.c} requires reading \path{audio/filter/af_scaletempo2_internals.h} to confirm operand types, while the path injection alert at \texttt{common/msg.c} requires following \path{root->stats_path} to \path{options/options.c} and \texttt{options/path.c}.
\textbf{Multi-turn interaction} is the second key capability: removing it reduces F1 to 56.3\%.
12 of the 14 full-correct cases lost by this ablation are FPs for \texttt{cpp/unused-static-variable}, where the one-shot variant sees only a declaration window but the full agent searches for later uses.
For example, the full agent follows \path{registry_listener} in \path{mpv@eadba44/player/clipboard/clipboard-wayland.c} from its declaration at line 272 to a use at line 336, and finds later reads of \path{g_ai2_ihevc_trans_32_13_815} in \path{libhevc@8cbcc58/common/arm/ihevc_resi_trans_neon_32x32.c}.
\textbf{Tool-based verification} has a smaller, localized effect, reducing F1 to 88.9\%.
The affected cases concentrate in integer-multiplication alerts, where the agent must reconcile typedefs, \texttt{sizeof}, and C integer-promotion semantics.
\textbf{Configuration-file access} has negligible impact: removing it loses no full-correct case, suggesting that these OSS-Fuzz C/C++ alerts depend mainly on source, header, macro, and type evidence rather than external configuration.

The baseline comparison further isolates the source of the agentic gain.
The LLM4SA-style prompt-template baseline improves over vanilla prompting, achieving 77.3\% FP recall and 70.8\% F1, but still falls short of the full agent.
The context-augmented baseline achieves 54.5\% FP recall and 66.7\% F1, showing that additional static context alone is insufficient.
Most importantly, the oracle-context baseline achieves only 36.4\% FP recall and 51.6\% F1, nearly identical to vanilla prompting despite receiving the same files accessed by \textsc{SWE-agent}.
This indicates that the advantage comes from iterative evidence gathering and reasoning, not merely from context volume.

\mynote{\textbf{Answer to RQ3.} On post-cutoff OSS-Fuzz C/C++ alerts, \textsc{SWE-agent} with Claude Sonnet 4 identifies 95.5\% of FPs with 95.5\% precision, versus 36.4\% FP identification for vanilla prompting. Ablations show that cross-file navigation and multi-turn interaction drive the gain, while context-only baselines remain below the full agent.}
}

\begin{table}[h]
\centering
\small
\caption{OSS-Fuzz C/C++ evaluation results. Metrics are percentages.}
\label{tab:ossfuzz_results}
\begin{tabular}{lcccccccc}
\toprule
Configuration & \makecell{Iden. \\ FP} & \makecell{Corr. \\ Ret.} & \makecell{Mis-id. \\ FP} & \makecell{Missed \\ FP} & Acc & Prec & Rec & F1 \\
\midrule
\cellcolor[HTML]{e1dcef}SWE-agent (full) & \cellcolor[HTML]{e1dcef}42.0 & \cellcolor[HTML]{e1dcef}54.0 & \cellcolor[HTML]{e1dcef}2.0 & \cellcolor[HTML]{e1dcef}2.0 & \cellcolor[HTML]{e1dcef}\underline{96.0} & \cellcolor[HTML]{e1dcef}\underline{95.5} & \cellcolor[HTML]{e1dcef}\textbf{95.5} & \cellcolor[HTML]{e1dcef}\textbf{95.5} \\
\quad w/o multi-turn & 18.0 & 52.0 & 2.0 & 26.0 & 70.0 & 90.0 & 40.9 & 56.3 \\
\cellcolor[HTML]{e1dcef}\quad w/o cross-file & \cellcolor[HTML]{e1dcef}20.0 & \cellcolor[HTML]{e1dcef}24.0 & \cellcolor[HTML]{e1dcef}0.0 & \cellcolor[HTML]{e1dcef}24.0 & \cellcolor[HTML]{e1dcef}44.0 & \cellcolor[HTML]{e1dcef}\textbf{100.0} & \cellcolor[HTML]{e1dcef}45.5 & \cellcolor[HTML]{e1dcef}62.5 \\
\quad w/o tool-based verification & 40.0 & 48.0 & 6.0 & 4.0 & 88.0 & 87.0 & \underline{90.9} & 88.9 \\
\cellcolor[HTML]{e1dcef}\quad w/o config read & \cellcolor[HTML]{e1dcef}42.0 & \cellcolor[HTML]{e1dcef}56.0 & \cellcolor[HTML]{e1dcef}0.0 & \cellcolor[HTML]{e1dcef}2.0 & \cellcolor[HTML]{e1dcef}\textbf{98.0} & \cellcolor[HTML]{e1dcef}\textbf{100.0} & \cellcolor[HTML]{e1dcef}\textbf{95.5} & \cellcolor[HTML]{e1dcef}\textbf{97.7} \\
Vanilla LLM & 16.0 & 52.0 & 2.0 & 28.0 & 68.0 & 88.9 & 36.4 & 51.6 \\
\cellcolor[HTML]{e1dcef}Vanilla LLM + LLM4SA & \cellcolor[HTML]{e1dcef}34.0 & \cellcolor[HTML]{e1dcef}38.0 & \cellcolor[HTML]{e1dcef}18.0 & \cellcolor[HTML]{e1dcef}10.0 & \cellcolor[HTML]{e1dcef}72.0 & \cellcolor[HTML]{e1dcef}65.4 & \cellcolor[HTML]{e1dcef} 77.3 & \cellcolor[HTML]{e1dcef}70.8 \\
Context-augmented Vanilla LLM & 24.0 & 52.0 & 4.0 & 20.0 & 76.0 & 85.7 & 54.5 & 66.7 \\
\cellcolor[HTML]{e1dcef}Vanilla LLM + Oracle Context & \cellcolor[HTML]{e1dcef}16.0 & \cellcolor[HTML]{e1dcef}54.0 & \cellcolor[HTML]{e1dcef}2.0 & \cellcolor[HTML]{e1dcef}28.0 & \cellcolor[HTML]{e1dcef}70.0 & \cellcolor[HTML]{e1dcef}88.9 & \cellcolor[HTML]{e1dcef}36.4 & \cellcolor[HTML]{e1dcef}51.6 \\
\bottomrule
\end{tabular}
\end{table}

\subsection{RQ\mrm{4}: Qualitative Analysis of Success and Failure Modes}

\begin{table*}[!th]
\centering

\begin{minipage}[t]{0.46\textwidth}\vspace{0pt}
    \centering
    \captionsetup{type=table,skip=2pt}
    \caption{Frequency of tool invocation across 256 analysis trajectories.}
    \label{tab:tool_use_freq}
    \begin{small} 
    \begin{tabular}{l r r r}
        \toprule
        Tool & \# Cases & \% Cases & Total Calls \\
        \midrule
        \texttt{view}       & 256 & 100.0 & 1,488 \\
        \texttt{find}       & 90  & 35.2  & 119 \\
        \texttt{grep}       & 82  & 32.0  & 118 \\
        \texttt{python3 -c} & 39  & 15.2  & 45 \\
        \texttt{javac/java} & 4   & 1.6   & 4 \\
        \bottomrule
    \end{tabular}
    \end{small}
\end{minipage}
\hfill
\begin{minipage}[t]{0.52\textwidth}\vspace{0pt}
    \centering
    \captionsetup{type=table,skip=2pt}
    \caption{Distribution of failure-mode (FM) patterns by CWE. A trajectory represents one agent run on one case; counts may exceed \#Traj. due to multiple patterns per trajectory.}
    \label{tab:failure_modes_by_cwe}
    \begin{small}
    \begin{tabular}{l r r r r}
        \toprule
        CWE & \# Traj. & FM1 & FM2 & FM3 \\
        \midrule
        CWE-327 & 85 & 85 & 35 & 0 \\
        CWE-614 & 24 & 15 & 18 & 0 \\
        CWE-79  & 8  & 0  & 0  & 8 \\
        \bottomrule
    \end{tabular}
    \end{small}
\end{minipage}

\end{table*}

\subsubsection{Capability Analysis: Tool Use and Reasoning}

Across the 256 successful trajectories, \textsc{SWE-agent} runs a median of 11 steps and views a median of 2 distinct files (source, config).
Notably, in 51.2\% of cases, the agent reads at least one non-target file beyond the primary benchmark test file, such as helper classes or configuration files.
In 25.0\% of cases, the final justification explicitly cites evidence found within these non-target files. 
Table~\ref{tab:tool_use_freq} shows a breakdown of how \textsc{SWE-agent} utilizes the tools.
We categorize the primary reasons for the agent's success into four patterns (P1--P4), as detailed below.

\noindent \textbf{P1: Cross-file Semantic Resolution.}
Many OWASP Benchmark FPs depend on semantics hidden in helper classes, e.g., wrapper sources, factory-selected implementations. 
Vanilla prompting an LLM without full context often overapproximates dataflow taint. 
For example, in \texttt{BenchmarkTest00866}, the vanilla LLM labels this instance as TP with high confidence, reasoning that a request parameter flows through a transformation and is concatenated into a file path (\texttt{TESTFILES\_DIR + bar}), thus enabling a \texttt{../} traversal.
However, the LLM's evidence is confined to the benchmark test file and assumes that the wrapper source is user-controlled.
In contrast, \textsc{SWE-agent} uses tool access to retrieve the missing non-local semantics.
It (1) opens \texttt{BenchmarkTest00866.java} to identify the dataflow, (2) inspects \texttt{SeparateClassRequest.java} and finds that \texttt{getTheValue()} is explicitly marked as a \emph{safe source} returning the constant string \texttt{"bar"} (hence the ``param'' is not attacker-controlled),
(3) inspects the factory-selected \texttt{Thing} implementation (\texttt{ThingFactory}/\texttt{Thing2} and \texttt{thing.properties}) to confirm the transformation does not reintroduce taint,
and (4) checks \texttt{Utils.TESTFILES\_DIR} to confirm a fixed safe base directory.
With these cross-file facts, the agent concludes that the effective filename is constant (base dir + \texttt{"bar"}), making traversal impossible, and correctly outputs \textsf{FP}.

\noindent \textbf{P2: Constant Folding for Control-flow Disambiguation.}
A common FP construction in the benchmark is a branch that looks data-dependent but is in fact constant.
Vanilla prompting frequently treats the unsafe branch as feasible, concluding TP.
\textsc{SWE-agent} often resolves this via a lightweight calculator call (e.g., \texttt{python3 -c}) and turns the control-flow argument into a verifiable numeric fact.
An example for this is \texttt{BenchmarkTest00105}, where the vanilla LLM flags SQL injection because \texttt{bar} \emph{could} be assigned from a cookie-derived \texttt{param}, which actually is a false branch in an unfold constant expression;
\textsc{SWE-agent} substitute the constant variable and computes that the constant\texttt{(7*18)+106 = 232 > 200} is always true, hence \texttt{bar} is always the constant \texttt{This\_should\_always\_happen} and the SQL query is not attacker-controlled.

\noindent \textbf{P3: Configuration Validation.}
For crypto and factory-driven behavior, the apparent default value in code can be misleading.
Vanilla prompting often assumes the fallback is vulnerable (e.g., default ECB mode), yielding a TP prediction.
\textsc{SWE-agent} resolves this by locating and reading the relevant \texttt{.properties} files, grounding the verdict in actual configuration.
This is exemplified by \texttt{BenchmarkTest01022}, where the vanilla LLM labels this instance as \textsf{TP} and cites two surface cues:
(1) a stack-trace print to the HTTP response, and
(2) a configuration call of the form \texttt{getProperty("cryptoAlg2", "AES/ECB/PKCS5Padding")}, which it interprets as using insecure ECB.

\textsc{SWE-agent} avoids this failure via configuration grounding and local semantic checks.
First, it inspects the relevant block and notes that the data being encrypted is not attacker-controlled due to a constant branch in \texttt{doSomething} (the condition is always true, forcing a constant string).
Second, rather than assuming the fallback cipher mode is used, it locates and opens src/main/resources/benchmark.properties and verifies that \texttt{cryptoAlg2} is set to \texttt{AES/CCM/NoPadding} (a non-ECB mode) in the benchmark configuration.
Therefore, the ``weak crypto'' alert is not supported under the benchmark execution context, and the agent outputs FP.

\noindent \textbf{P4: Direct Verification of Complex Semantics.}
In a small number of cases (4/256, 1.6\%), the agent attempts an even stronger form of grounding: it generates a minimal Java snippet that reproduces the suspicious logic and tries to compile/run it (e.g., list index manipulation).
For example, in \texttt{BenchmarkTest00265}, the vanilla LLM treats a list-based transformation as propagating user input into a file path.
\textsc{SWE-agent} creates a miniature program (\texttt{test\_logic.java}) to validate the effect of \texttt{add/remove/get} on the selected element.
In our sandbox, \texttt{javac} is unavailable, so execution does not succeed; nevertheless, this trajectory suggests that enabling lightweight execution-based checks when feasible could further improve on semantics analysis that are easy to misread but cheap to validate.
\subsubsection{Failure Analysis}
The inspection of 103 failed trajectories reveals three primary failure modes (FMs).
Table~\ref{tab:failure_modes_by_cwe} presents the distribution of these FMs.

\noindent \textbf{FM1. Incorrect CWE Attribution.} 
Agents often validate a security issue other than the benchmark's target CWE. In all CWE-327 cases, the source code contains \path{printStackTrace(response.getWriter())} within a cryptographic exception handler. This pattern may reveal sensitive application information to users. The trajectories indicate that agents identify this as a true vulnerability related to the cryptographic context, even though it does not constitute a CWE-327 violation. Furthermore, 7 out of 10 CWE-614 cases are marked as TP because the agents conclude that the cookie value is user-controlled and susceptible to CRLF injection (CWE-113) due to a lack of sanitization, rather than focusing on the missing Secure flag.


\noindent \textbf{FM2. Overly Conservative Threat Modeling.} 
Agents frequently adopt a worst-case threat model, which prioritizes recall but reduces precision in FP triage. 
Even when agents focus on the SAST alerts, they often conflate hardening guideline violations with exploitable vulnerabilities. 
In 4 out of 10 CWE-614 cases, agents report a TP solely because the \texttt{SameSite} attribute is absent. 
The agents reason that this absence could permit a CSRF attack despite hardcoded test inputs, arguing that the code pattern represents a production security weakness without attempting to construct a feasible payload. 
Additionally, in 12 out of 47 CWE-327 cases, agents identify weak cryptographic algorithms such as \texttt{"AES/ECB/PKCS5Padding"} that are unreachable in the execution flow. The reasoning follows a static-like pattern: the existence of the weak algorithm justifies reporting a vulnerability.


\noindent \textbf{FM3. Surface-Level Pattern Matching.}
Agents may rely on surface cues, such as specific API names and suspicious sinks, and terminate analysis prematurely without deep semantic validation. In all 4 CWE-79 cases, both agents conclude a TP based on the presence of \texttt{format()} and output APIs. This combination is often associated with format-string or injection vulnerabilities. However, the data returned in these cases is already HTML-escaped. The remaining concern involves software robustness rather than a functional XSS primitive.

\mynote{
\textbf{Answer to RQ\mrm{4}:}
Agentic workflows improve FP filtering through cross-file semantic resolution and control-flow disambiguation, with 51.2\% of successful cases relying on evidence from non-target files such as helper classes and configurations.
However, performance is limited by conservative threat modeling and incorrect CWE attribution, where agents often prioritize hardening guidelines or side-channel leaks over the specific benchmark ground truth.
}

\section{Discussion}
\label{sec:discussion}

\subsection{Cost Analysis} 
\begin{table*}[!th]
\centering

\begin{minipage}[t]{0.55\textwidth}\vspace{0pt}
  \centering
  \captionsetup{type=table,skip=2pt}
  \caption{Average resource consumption and operational metrics per task across different models and agents.}
  \label{tab:agent_cost_analysis}
  \begin{adjustbox}{max width=\linewidth}
    \begin{tabular}{llrrr}
      \toprule
      Model & Agent & Avg. Rounds & Avg. Tokens & Avg. Cost (USD) \\
      \midrule
      \multirow{3}{*}{Claude Sonnet 4}
        & \textsc{Aider}     & 1.0  & 13,321  & 0.0469 \\
        & \textsc{OpenHands} & 20.5 & 219,299 & 0.1867 \\
        & \textsc{SWE-agent} & 9.0  & 130,846 & 0.1501 \\
      \midrule
      \multirow{3}{*}{DeepSeek Chat}
        & \textsc{Aider}     & 1.0  & 11,036  & 0.0028 \\
        & \textsc{OpenHands} & 27.1 & 269,260 & 0.0252 \\
        & \textsc{SWE-agent} & 13.6 & 189,937 & 0.0215 \\
      \midrule
      \multirow{3}{*}{GPT-5}
        & \textsc{Aider}     & 1.0  & 12,564  & 0.0311 \\
        & \textsc{OpenHands} & 7.4  & 75,205  & 0.0599 \\
        & \textsc{SWE-agent} & 8.3  & 125,106 & 0.1052 \\
      \bottomrule
    \end{tabular}
  \end{adjustbox}
\end{minipage}\hfill
\begin{minipage}[t]{0.42\textwidth}\vspace{0pt}
  \centering
  \includegraphics[width=0.92\linewidth]{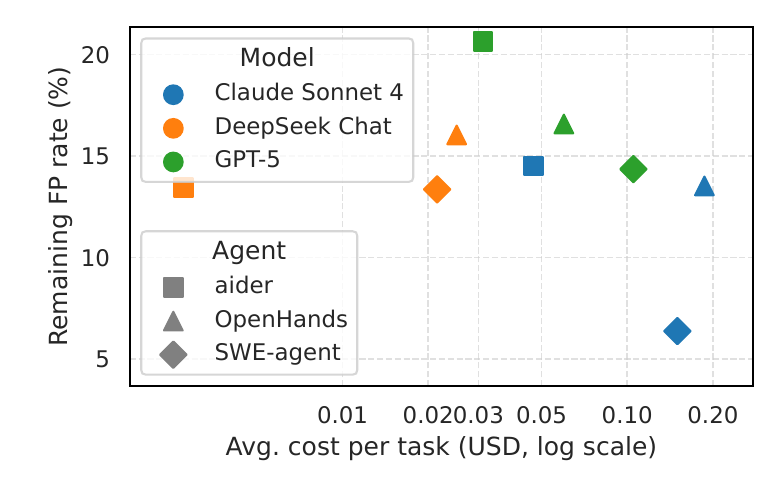}
  \captionsetup{type=figure,skip=2pt}
  \caption{Cost effectiveness of agent frameworks (OWASP FP Reduction). Lower left is better.}
  \label{fig:cost_vs_remaining_fp_rate_owasp}
\end{minipage}

\vspace{-2mm}
\end{table*}

\looseness=-1
Table~\ref{tab:agent_cost_analysis} and Figure~\ref{fig:cost_vs_remaining_fp_rate_owasp} show that agent frameworks differ sharply in interaction cost. \textsc{OpenHands} is the most iterative, averaging 20.5 rounds with Claude Sonnet 4 and 27.1 with DeepSeek Chat, while \textsc{Aider} finishes in one round and behaves closer to a prompting engine. \textsc{SWE-agent} uses fewer rounds than \textsc{OpenHands} but much heavier context per round, consuming 130,846 tokens per Claude run. Cost follows these patterns: \textsc{OpenHands}+Claude is the most expensive per task (\$0.1867), \textsc{SWE-agent}+GPT-5 is costly due to token load (\$0.1052), and \textsc{Aider}+DeepSeek is the cheapest (\$0.0028). These results show a practical cost--effectiveness frontier with no uniformly dominant agent.

\subsection{True-Positive Retention}
\label{sec:tp-retention}
\looseness=-1
While RQ1 focuses on filtering FPs among SAST alerts, an equally important concern is whether these agents suppress true vulnerabilities. To quantify this risk, we conducted a TP-retention experiment on the OWASP Benchmark positives using the best-performing configuration from Table~\ref{tab:r1_fp_reduction_performance}, i.e., Claude Sonnet 4 paired with \textsc{SWE-agent}. Among the 1,415 vulnerable test cases, the agent successfully completed 1,411 runs, achieving a 99.7\% completion rate.
The agent incorrectly labeled 314 true vulnerabilities as FPs, resulting in a TP retention of 77.7\% and a miss rate of 22.25\%.
These results suggest that despite strong FP reduction on non-vulnerable instances, the agent may still hide a non-trivial fraction of real vulnerabilities if its decisions drive automatic suppression.

Our per-CWE analysis shows that the miss rate remains low for classic data-flow-driven injection vulnerabilities. Specifically, the miss rate is 2.38\% for CWE-78, 1.10\% for CWE-89, 0.41\% for CWE-79, and 0\% for both CWE-90 and CWE-643. 
This indicates that the agent is generally reliable when reasoning about explicit taint-flow evidence in local code. In contrast, we observe severe TP suppression for weaknesses that are less obvious to straightforward data-flow reasoning. For instance, the miss rate rises to 77.17\% for CWE-327 (Weak Cryptography), 84.50\% for CWE-328 (Weak Hashing), 77.11\% for CWE-501 (Trust Boundary Violation), and 50.00\% for CWE-614 (Secure Cookie Flag). These categories often require domain knowledge such as algorithm strength, configuration context, or threat modeling. In these cases, the agent may dismiss alerts as best-practice warnings rather than exploitable vulnerabilities, or infer mitigations that do not exist.

\mr{R4}{
\subsection{Data Contamination Check}
A potential concern is that the models may have seen our evaluation datasets during training (data leakage), inflating performance through memorization rather than genuine reasoning capability.
We address this through three complementary checks.
First, we conducted a filename-only leakage probe on the OWASP Benchmark: we provided only the case identifier (e.g., \texttt{BenchmarkTest00171}), then asked each model to predict whether the case was vulnerable or non-vulnerable. All three models remained at chance level (Claude Sonnet 4: 51.6\% accuracy, GPT-5: 51.0\%, DeepSeek Chat: 49.8\%), indicating no label memorization.
Second, we scanned all 15,636 OWASP agent trajectories for explicit prior-exposure language audit such as ``training data,'' ``memorized,'' or ``I've seen this before.'' No trajectory contained claims of prior benchmark, file, or answer exposure.
Third, we conducted an evaluation on five OSS-Fuzz projects: mpv, libhevc, libical, gpsd, and quickjs. We selected them by scanning backward from the most recent OSS-Fuzz records in the OSV export, deriving vulnerable commits from OSV fixing commits, constructing vulnerable repository snapshots, and running CodeQL. The selected vulnerable and fixing commits postdate the evaluated model's known training cutoff by 10--12 months. On this contamination-free C/C++ dataset, the agent still achieves 95.5\% F1, indicating that the gains generalize beyond public Java benchmarks.
}


\mr{R5}{
\subsection{Scope Drift in Weaker Models}
We first observe a counterintuitive degradation for DeepSeek Chat in RQ1: the vanilla baseline achieves the lowest remaining FPR among DeepSeek configurations, whereas adding \textsc{SWE-agent} increases it.
To understand why, we conduct a targeted trajectory audit rather than attributing the result to generic tool-use noise.
The audit compares 1,303 paired OWASP runs for which both vanilla DeepSeek and DeepSeek + \textsc{SWE-agent} completed.
We identify 145 cases where vanilla DeepSeek is correct but DeepSeek + \textsc{SWE-agent} is wrong.
All 145 are ground-truth non-vulnerable cases, and in every case the agent changes the classification to TP.
Thus, the degradation is a precision failure: additional agentic behaviour causes DeepSeek to over-report FPs as vulnerabilities.
We then audit the final trajectory rationales for these 145 cases; the categories overlap, but the pattern is clear: 80 trajectories mention stack-trace or \texttt{printStackTrace} exposure, 93 mention information or error disclosure, 54 mention cookie-hardening issues such as \texttt{HttpOnly} or \texttt{SameSite}, 23 mention carriage-return/line-feed or response-splitting concerns, 35 mention weak cryptography, and 97 mention exception-handling concerns such as \texttt{IOException}.
Three example cases illustrate this scope drift.
In \texttt{BenchmarkTest00054}, vanilla DeepSeek correctly returns an FP verdict because the cross-site scripting path is HTML-encoded, whereas \textsc{SWE-agent} returns a TP verdict by aggregating stack-trace exposure, missing cookie attributes, and unhandled \texttt{IOException}.
In \texttt{BenchmarkTest00089}, the agent converts a secure-cookie FP verdict into a speculative carriage-return/line-feed response-splitting TP verdict.
In \texttt{BenchmarkTest00225}, it escalates ordinary servlet exception propagation into resource-leak, denial-of-service, and information-disclosure concerns.
These cases lead to a scope-drift explanation: for DeepSeek, the agentic loop shifts from alert verification to broader vulnerability hunting.
The extra context therefore amplifies speculative hypotheses instead of correcting them, degrading precision despite increased exploration.
}

\subsection{Lessons Learnt}
\noindent\textbf{Lesson 1: LLM-based agents can remove most SAST FPs, but only for certain vulnerability categories.}
\textit{Evidence.} Section~\ref{rq1:result} shows that LLM-based agents can reduce SAST noise on the OWASP Benchmark: the best configuration, i.e., Claude Sonnet 4 with \textsc{SWE-agent}, lowers the remaining FPR from over 98\% to 6.3\%, eliminating more than 92\% of FPs (Table~\ref{tab:r1_fp_reduction_performance}).
However, the CWE-level analysis in Section~\ref{sec:categorical-analysis-across-cwes} reveals a strong skew: residual FPs are concentrated in a few categories, particularly CWE-327, CWE-330, and CWE-614, while injection-style vulnerabilities (e.g., CWE-78, CWE-79, CWE-89) are filtered almost completely. \textit{Lesson.}
LLM-based agents are highly effective FP filters for data-flow-driven vulnerabilities, but their effectiveness drops sharply for policy- and cryptography-related weaknesses that require domain knowledge or threat modeling.

\noindent\textbf{Lesson 2: The value of agentic reasoning depends more on the backbone model than on the agent framework.}
\looseness=-1
\textit{Evidence.} The comparison in Section~\ref{rq1:result} shows that agentic frameworks consistently outperform vanilla LLM prompting for stronger backbone models such as Claude Sonnet 4 and GPT-5, reducing the remaining FPR from 23.0\% to 6.3\% for Claude.
In contrast, for DeepSeek Chat, vanilla prompting already achieves a low remaining FPR (11.2\%), and adding agentic scaffolding provides no consistent benefit and can even degrade performance.
This backbone dependency is further confirmed by the per-CWE agentic behaviour gain/loss analysis in Section~\ref{sec:categorical-analysis-across-cwes} (Figure~\ref{fig:agentic_gain_heatmap}). \textit{Lesson.}
Agentic reasoning amplifies the strengths of capable backbone models but does not compensate for weaker ones; backbone selection matters more than agent framework choice.

\noindent\textbf{Lesson 3: Aggressive FP suppression risks hiding real vulnerabilities and should not be fully automated.}
\textit{Evidence.}
While Section~\ref{sec:results} focuses on FP reduction on non-vulnerable cases, the TP retention analysis in Section~\ref{sec:tp-retention} shows that even the best-performing configuration incorrectly suppresses 22.25\% of real vulnerabilities on the OWASP Benchmark positives.
This risk is highly CWE-dependent: miss rates are negligible for injection-style vulnerabilities but exceed 50\% for cryptography- and policy-related categories such as CWE-327, CWE-328, CWE-501, and CWE-614. \textit{Lesson.}
LLM-based agents are unsuitable for unconditional, automatic suppression of SAST warnings; they should instead be deployed as decision-support tools, especially for vulnerability categories with high safety risk.

\subsection{Threats to Validity}

\noindent\textbf{Threats to Internal Validity.}
\looseness=-1
The first threat stems from the inherent non-determinism of LLMs, which may compound across multiple reasoning steps in agentic frameworks and affect the reproducibility of this study.
We mitigate this by setting the temperature to zero, which promotes, but does not guarantee, deterministic results \cite{astekin2024exploratory}.
\mrm{Another threat concerns the size and labeling process of the real-world alert samples.
For Vul4J, our RQ2 evaluation includes 50 sampled CodeQL alerts due to limited manual triage capacity, and most labels were assigned by the first author after patch matching.}
Although we use a fixed random seed to ensure reproducibility, the sample may not fully represent the original alert distribution, and the measured performance can be affected by the rule mix (some CodeQL rules are inherently noisier than others).
\mrm{For OSS-Fuzz, we mitigate this concern through an independent dual-rater protocol with substantial agreement, but the dataset is still a focused 50-alert validation set rather than a comprehensive C/C++ benchmark.}
\mr{R7}{
Also, our evaluation follows prior work on static-analysis warning validation, where the reported static-analysis output (alert) is the unit of analysis and each warning is judged against an alert-specific TP/FP label~\cite{kang2022detecting,joshy2021validating,zheng2021d2a}.
This design matches the SAST triage setting: developers must decide whether to act on the reported alert, not whether the surrounding code contains any possible security issue.
However, this alert-specific framing has a limitation: an agent may correctly reject the reported vulnerability rationale while identifying a different real security issue in the same code path.
Under our current metric definition, such cases count as incorrect because correctness is defined with respect to the specific labeled alert rather than broader security utility.
We acknowledge that this binary framing may undercount cases where the agent provides actionable security value beyond the scope of the original alert.
}

\noindent\textbf{Threats to External Validity.}
\looseness=-1
To enhance the generalizability of our findings, \mrm{our evaluation combines a controlled Java benchmark, real-world Java projects, and a focused post-cutoff C/C++ validation set.
However, the full cross-model and cross-agent comparison is conducted on OWASP and Vul4J, while OSS-Fuzz is used specifically to test contamination-free generalization and isolate agentic capabilities for the strongest configuration.
Therefore, the OSS-Fuzz results should not be interpreted as a complete ranking of all agents or backbone models on C/C++.}
Second, although our real-world evaluation primarily utilizes CodeQL alerts, we mitigate tool-specific bias by first analyzing the correlation of FPs across multiple industrial scanners (Semgrep and SonarQube), ensuring that our agent systems target shared semantic blind spots rather than isolated analyzer errors. Finally, we evaluate three state-of-the-art models and agent frameworks to reduce selection bias.
\section{Conclusion and Future Work}

In this study, we provide a comparative evaluation of three LLM-based agent frameworks, i.e., \textsc{Aider}, \textsc{OpenHands}, and \textsc{SWE-agent}, for filtering FPs generated by SAST tools.
Experimental results indicate that agentic reasoning is effective. 
In the best-performing configuration, using \textsc{SWE-agent} with Claude Sonnet 4, the agentic workflow reduced the initial FP rate of 98.3\% on the OWASP Benchmark to 6.3\%, effectively removing over 92\% of the noise.
Similarly, in real-world scenarios involving CodeQL alerts \mrm{on Java}, the agents achieved an FP identification rate of up to 93.3\%. 
Further evaluation on unseen C/C++ vulnerabilities sampled from OSS-Fuzz shows that this advantage is not explained solely by memorization of public Java benchmarks: on these projects, \textsc{SWE-agent} with Claude Sonnet 4 achieves 96.0\% accuracy and 95.5\% F1, substantially outperforming vanilla, LLM4SA-style, context-augmented, and oracle-context baselines. 
The effectiveness of these agents is non-uniform and depends on both the backbone model and the vulnerability category. 
Claude Sonnet 4 and GPT-5 improved filtering performance when deployed within an agentic framework compared to vanilla zero-shot prompting, reducing residual FPR from 23.0\% to 6.3\%, whereas weaker backbones showed no consistent gain from added agentic loops.
However, this reliability is concentrated in data-flow-dependent injection categories, whereas weakness families that require domain-specific heuristics or policy understanding retain higher residual FPs and carry a substantially greater risk of suppressing true positives.

In the future, we plan to develop CWE-aware, cost-adaptive agent strategies and human-in-the-loop auditing mechanisms to maintain high FP filtering rates while minimizing the loss of TPs, and to evaluate their generalizability across additional programming languages and SAST tools.


\newpage
\section*{Data-Availability Statement}

To ensure the reproducibility of our results and to provide transparency in our research, we have made all related scripts and data publicly available. All resources can be accessed as part of our anonymized artifact, which is available at \href{https://doi.org/10.5281/zenodo.18420284}{https://doi.org/10.5281/zenodo.18420284}~\cite{anonymous_2026_21282004}.

\bibliographystyle{ACM-Reference-Format}
\bibliography{main}

\end{document}